\documentclass[preprint,showpacs,aps]{revtex4}
\usepackage{epsfig}

\begin{document}
\title{Voltage switching and domain relocation in semiconductor superlattices}

\author{ L. L. Bonilla\cite{bonilla:email}, R. Escobedo and G. Dell'Acqua }

\affiliation{Grupo de Modelizaci\'on y Simulaci\'on Num\'erica,  
Universidad Carlos III de Madrid, Avenida de la Universidad 30, 28911 Legan{\'e}s, Spain}

\date{ \today  }

\begin{abstract}
A numerical study of domain wall relocation during voltage switching with different ramping
times is presented for weakly coupled, doped semiconductor superlattices exhibiting 
multistable domain formation in the first plateau of their current-voltage characteristics. 
Stable self-oscillations of the current at the end of stable stationary 
branches of the current-voltage characteristics have been found. These oscillations are due to 
periodic motion of charge dipoles near the cathode that disappear inside the SL, before they 
can reach the receiving contact. Depending on the dc voltage step, the type of multistability 
between static branches and the duration of voltage switching, unusual relocation scenarios 
are found including changes of the current that follow adiabatically the stable I--V branches, 
different faster episodes involving charge tripoles and dipoles, and even small amplitude 
oscillations of the current near the end of static I--V branches followed by dipole-tripole
scenarios. 
\end{abstract}

\keywords{superlattices, relocation of domains, voltage switching}
\pacs{73.63.-b, 05.45.-a, 73.23.-b, 73.40.-c}

\maketitle

\section{Introduction}
\label{sec:1}
Nonlinear vertical electron transport in semiconductor superlattices (SL) gives rise to a rich 
variety of dynamical phenomena associated with negative differential conductivity (NDC)
\cite{Wac02,Bon02,PAg04,BGr05}. In wide miniband SL, NDC due to Bragg scattering is 
the origin of self-oscillations of the current through a dc voltage biased SL 
\cite{BTh77,HGS96}. These oscillations are due to recycling of charge dipoles as in the 
Gunn effect of bulk GaAs. In weakly coupled SL, NDC due to sequential tunneling between 
quantum wells may cause either stable self-sustained current oscillations mediated by traveling 
charge monopoles or dipoles \cite{KKG95,KHP97,SMB99}, or a sawtooth multistable 
current--voltage (I--V) characteristics associated with static field domains 
\cite{ECh74,KWA86,GKP91}. The observed behavior depends crucially on the SL
configuration (widths of wells and barriers, number of SL periods, boundary conditions in the
contact regions), doping density and voltage bias \cite{BGr05}. 

Although there remain important gaps in our theoretical knowledge of nonlinear transport
in SL \cite{Wac02,BGr05}, a basic difference between weakly and strongly coupled SL
is the type of balance equations describing them. Weakly coupled SL are described by
spatially discrete balance equations whereas spatially continuous balance equations 
determine the nonlinear behavior of strongly coupled SL \cite{Bon02}. Traces of 
spatial discreteness are the sawtooth I--V characteristics and the current spikes during
self-oscillations and during domain relocation due to voltage switching; see the review of
theory and experiments in \cite{BGr05}. 

Recently, the dynamic response of the current and the field profile to voltage switching has 
been investigated both experimentally \cite{REM97,LGP98,RTG01,RTG02,RGT02} and 
theoretically \cite{AWB01,BSS02}. For a SL with a multistable I--V characteristics, each 
branch thereof corresponds to having the domain wall separating the low and high field 
domains of the field profile (which is a charge accumulation layer, CAL) placed at a different 
well of the SL. We are interested in the transition from one stable stationary branch of the 
I--V characteristics to another due to a step in the applied voltage $\Delta V = V_{f}-
V_{i}$ ($V_{i}$ and $V_{f}$ are the initial and final voltage values). Bias steps are turned 
on during a short time interval called {\em ramping time}, which can be zero. A bias step 
increasing the applied dc voltage is referred to as an {\em up jump} ($\Delta V>0$), while a 
bias step decreasing the applied dc voltage is called a {\em down jump} ($\Delta V<0$). 
Bias steps contrast with voltage up-sweeps and down-sweeps for which the ramping time is
infinitely long and the bias increase or decrease is adiabatic. For down 
jumps, the relocation process of the domain boundary proceeds via a direct motion of the 
CAL in the direction of electron flow. This behavior is confirmed by single-shot time traces 
of the current response: for values of $V_{f}$ away from regions of bistability, there is an 
initial displacement current spike, after which the current rapidly switches to the stable value. 
Furthermore, when $V_{f}$ is near to the bistable region, there is an additional intermediate 
period, in which the current fluctuates about a metastable value for a stochastically varying 
delay time $\tau_d$, before rapidly switching to the stable value in a time $\tau_s$ 
\cite{LGP98,RTG01}. However, for up jumps, the charge monopole at the domain 
boundary would have to move against the electron flow, and this is only possible for 
small-amplitude up jumps. For larger up jumps, the more complex dipole-tripole scenario 
occurs: the CAL moves one well against the electron flow, then a charge dipole comprising 
one CAL and one charge depletion layer (CDL) is formed at the cathode and, together with 
the old CAL, it moves with the electron flow. The resulting charge tripole exists until the old 
CAL reaches the anode and disappears, leaving only the charge dipole. The CDL of this 
dipole reaches the anode while its CAL moves until its final position at the destination branch 
of the I--V characteristics \cite{AWB01,RTG02}. For $V_{f}$ near the end of a branch, 
there are pronounced stochastic effects due to shot noise \cite{RTG01,RGT02}.

In this paper, we carry out an extensive numerical study of the dynamical response to voltage 
switching and unveil unexpected behavior. We use a discrete sequential tunneling model 
whose detailed description can be found in Appendix A of \cite{Bon02}. Stochastic effects 
due to shot noise will be ignored. We find stable self-oscillations of
the current at the end of stable stationary branches of the I--V characteristics. These
oscillations are due to periodic motion of charge dipoles near the cathode that disappear
inside the SL, before they can reach the receiving contact. We also want to understand
how the dynamical response to voltage switching is affected by the number of multistable 
branches of the I--V characteristics, their extension and the {\em ramping time} necessary to 
change voltage (from $V_{i}$ to $V_{f}$). Among our results, we 
find a different tripole-dipole scenario than that reported by Amann et al \cite{AWB01}
(the first phase of the scenario is different). We also find that the ramping time selects the 
tripole--dipole scenario for large up jumps. Suppose that there are several branches of the 
I--V characteristics in the interval between $V_{i}$ and $V_{f}$ (large voltage switching).
Then there are two critical ramping times $\tau_{c1}$ and $\tau_{c2}$, $\tau_{c2}<
\tau_{c1}$, whose precise values depend on the SL parameters in Table \ref{SLvalues} 
and on $V_{i}$ and $V_{f}$. For the parameters used in our simulations, the critical ramping
times are between 10 and 30 $\mu$s. If the ramping time is larger than $\tau_{c1}$, the 
current follows adiabatically the stable branches until their end, falls to the next stable branch 
and repeats this process until $V_{f}$ is reached. For ramping times between $\tau_{c2}$ 
and $\tau_{c1}$, adiabatic motion over a stable branch is followed by a tripole--dipole 
scenario until the following stable branch is reached. Depending on the number of multistable 
branches, sometimes a stable branch is skipped in this process. Lastly, if the ramping time is 
shorter than $\tau_{c2}$, the final stable branch is reached after only one tripole--dipole 
scenario even for large voltage steps. 

The rest of the paper is as follows. The model we use and details of its numerical integration
are described in Sections \ref{sec:2} and \ref{sec:3}, respectively. Section \ref{sec:4}
contains the multistable I--V characteristics of a SL with realistic configuration and doping 
density parameters \cite{KHP97}. We show that the width of the multistability regions
increases with voltage while the slope of the branches (which is the positive differential 
conductivity or, in short, PDC) decreases. These features affect substantially the dynamic 
response to switching described in Sections \ref{sec:5} and \ref{sec:6}. Finally, the main 
results of the paper are summarized in Section \ref{sec:7}. 
 
\section{Sequential tunneling discrete model}
\label{sec:2}
In weakly coupled SL, typically the scattering times (about 0.1 ps) are much shorter than
the escape times from quantum wells (about 0.01 ns for a SL with a 0.1 meV miniband
width). In their turn, the latter are shorter than typical dielectric relaxation times (on the
nanosecond time scale) \cite{BGr05}. This implies that the dominant mechanism of vertical 
charge transport is sequential resonant tunneling and that the tunneling current across barriers 
can be considered to be stationary on the time scale of dielectric relaxation. Nonlinear 
stationary and oscillatory phenomena occurring for voltages in the first plateau of weakly 
coupled doped SL have been well described by the spatially discrete model equations (with 
backward finite differences) introduced in \cite{Bon95} with constitutive relations between 
sequential tunneling current, electron densities and electric field of the type calculated in 
\cite{Wac98} using stationary nonequilibrium Green functions or in \cite{Bon02} (and
references cited therein) approximating Transfer Hamiltonian formulas. See the review 
\cite{BGr05} for a recent description and further justification. The model equations consist
of the Poisson and charge continuity equations for the two-dimensional (2D) electron
density $n_i$ and average electric field $-F_i$ at the $i$th SL period (which starts at the
right end of the ($i-1$)th barrier and finishes at the right end of the $i$th barrier)
\begin{eqnarray}
&& F_i - F_{i-1} = \frac e{\varepsilon} (n_i - N_D), \label{e1}\\
&& \frac{{\rm d}n_i}{{\rm d}t} = J_{i-1\rightarrow i} - J_{i\rightarrow i+1} ,
\quad  \, i=1, \cdots, N .  \label{e2}
\end{eqnarray}
Here $N_D$, $\varepsilon$, $-e$ and $eJ_{i\rightarrow i+1}$ are the 2D doping density 
at the $i$th well, the average permittivity, the electron charge and the tunnelling
current density across the $i$th barrier, respectively. The width of a SL period
is $l=d+w$, where $d$ and $w$ are the barrier and well widths, respectively.
Time-differencing Eq.\ (\ref{e1}) and inserting the result in Eq.~(\ref{e2}),
we obtain the following form of Ampere's law,
\begin{equation}
\varepsilon\, \frac{{\rm d}F_i}{{\rm d}t} + J_{i\rightarrow i+1} = J(t),
\label{e3}
\end{equation}
which may be solved with the bias condition for the applied voltage $V(t)$:
\begin{eqnarray}
{1 \over N+1} \sum_{i=0}^N F_i = {V(t) \over (N+1)\, l}.
\label{e4-bias}
\end{eqnarray}
The space-independent unknown function $J(t)$ is the total current density through the SL. 
The $2N+2$ independent equations of the discrete model are (\ref{e1}) for $i=1,\ldots,N$, 
(\ref{e3}) for $i=0,\ldots, N$ and (\ref{e4-bias}) for the $2N+2$ unknowns $n_{i}$, 
$F_{i}$ ($i=1,\ldots,N$), $F_{0}$ and $J$, provided we have $N+1$ constitutive relations 
linking the tunneling current $J_{i\to i+1}$ ($i=0,\ldots,N$) to the electron densities and
electric fields. To calculate the tunneling currents across SL barriers, we use explicit formulas 
provided by the Transfer Hamiltonian method when the scattering broadening is much smaller 
than the subband energies and chemical potentials \cite{Bon02}: 
\begin{eqnarray}
J_{i\to i+1} & = & {e\, v^{(f)} (F_i) \over l} \left\{n_i - {m^* k_B T\over\pi
\hbar^2} \ln \left[ 1 + \exp \left( - {e F_i l \over k_B T} \right) \right.  \right.
\nonumber \\
&& \left. \left.
\times \left( \exp \left({\pi \hbar^2 n_{i+1} \over m^* k_B T} \right) -1
 \right)\right] \right\}, \quad i=1,\dots,N-1, \label{Jtuni}
\\
J_{0\to 1} & = & \sigma F_0, \label{Jtun0}
\\
J_{N\to N+1} & = & \sigma F_N {n_N \over N_D}.
\label{JtunN}
\end{eqnarray}
As boundary tunneling currents for $i=0$ and $N$, we adopt linear relations between current 
and field as in Ref.~\cite{AWB01}. In these formulas, $\sigma$ is the contact conductivity 
(assumed to be the same at both contacts for simplicity), $m^*$ the effective mass, $T$ the
temperature, $k_B$ and $\hbar$ are the Boltzmann and Planck constants respectively, and 
the ``forward tunneling velocity'' $v^{(f)}$ is a sum of Lorentzians centered at the resonant
field values $F_{C\nu}= (\mathcal{E}_{C\nu}-\mathcal{E}_{C1})/(el)$:
\begin{eqnarray}
v^{(f)} (F_i) & = & \sum_{j=1}^n
\frac{ {\hbar^3 l (\gamma_{C1} + \gamma_{C_j}) \over 2m^{*2}}\,
{\cal T}_i ({\cal E}_{C1}) }
{ ({\cal E}_{C1} - {\cal E}_{C_j} + eF_il)^2 + (\gamma_{C1} + \gamma_{C_j})^2},
\label{superVEL}\\
{\cal T}_i (\epsilon) & = & \frac{16 k_i^2 k_{i+1}^2 \alpha_i^2
 (k_i^2 + \alpha_i^2)^{-1} (k_{i+1}^2 + \alpha_i^2)^{-1} }
{(w + \alpha_{i-1}^{-1} + \alpha_i^{-1}) (w + \alpha_{i+1}^{-1} + \alpha_i^{-1})
 e^{2 \alpha_i d} },
\\
\hbar k_i & = & \sqrt{2 m^* \epsilon}, \\
\hbar k_{i+1} & = & \sqrt{2 m^* [\epsilon + e (d+w) F_i]}, \\
\hbar \alpha_{i-1} & = & \sqrt{2 m^* \left[ e V_b + e \left( d + {w\over 2} \right) F_i
- \epsilon )\right]}, \\
\hbar \alpha_i & = & \sqrt{2 m^* \left[ e V_b - {e w F_i \over 2} - \epsilon )\right]}, \\
\hbar \alpha_{i+1} & = & \sqrt{2 m^* \left[ e V_b - e \left( d + {3w\over 2} \right) F_i
- \epsilon )\right]}. \label{last}
\end{eqnarray}
In these equations, $C_j$ indicates the $j$th subband in a well, ${\cal E}_{C_j}$
is its energy measured from the bottom of the well, $\gamma_{C_j}$ is the scattering width, 
${\cal T}_i$ is a dimensionless transmission probability across the $i$th barrier,
and $eV_b$ is the barrier height in absence of potential drops. Typical
values of these parameters are shown in Table~\ref{SLvalues}.

\begin{table}[ht]
\caption{Parameters of the 9/4 SL in Ref~\cite{KHP97}.}
\begin{center} \footnotesize
\begin{tabular}{cccccccccc}
 \hline
$N$ & $N_D$ & $d_{W}/d_{B}$ & $\gamma$ & $m^*$ &
 ${\cal E}_{C_1}$ & ${\cal E}_{C_2}$ & ${\cal E}_{C_3}$ & $V_b$ \\
& (cm$^{-2}$) & (nm/nm) & (meV) & ($10^{-32}$ Kg) & (meV) & (meV) & (meV) & (V) \\\\
40 & $1.5 \times 10^{11}$ & 9.0/4.0 & 8 & $8.43$ & 44 & 180 & 410 & 0.982  \\
 \hline
\end{tabular}
\end{center}
\label{SLvalues}
\vspace{-0.4cm}
\end{table}

To carry out numerical integrations of the discrete model, the explicit formulas (\ref{Jtuni}) 
- (\ref{last}) are much better than numerically calculated tunneling currents such as those
obtained in \cite{Wac98,Wac02} from Green function calculations. Furthermore, explicit
tunneling currents are better suited for analysis of the discrete model equations. These reasons
to favor the previous explicit formulas for the sequential tunneling current are not offset by 
claims that one type of derivation (Transfer Hamiltonian or Green functions) agrees better
with first principles: both derivations involve similarly drastic simplifications and the 
resulting formulas agree similarly well with experiments. It turns out that the type of
solutions of the discrete model depends on the {\em qualitative features} of $J_{i\to i+1}$
as a function of $F_{i}$, $n_{i}$ and $n_{i+1}$, not on detailed quantitative features. The
formulas for $J_{i\to i+1}$ obtained using different derivation methods yield a 
tunneling current similar to that in Fig.~\ref{fig1}, which is why we obtain similar results.
See discussions in Ref.~\cite{BGr05}. 

\section{Nondimensionalization and numerical integration}
\label{sec:3}
\begin{figure}[ht]
\begin{center}
\epsfxsize=100mm
\epsfbox{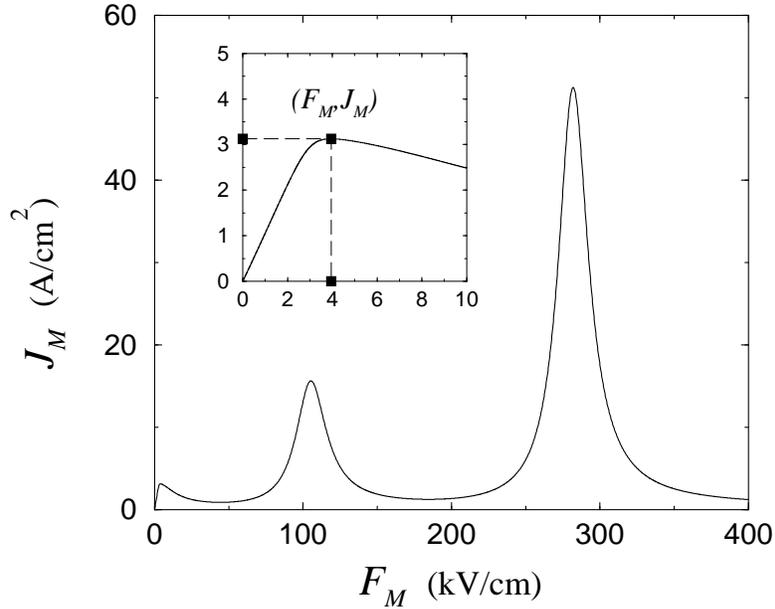}
\caption{Tunneling current density for $n_{i}=n_{i+1}=N_{D}$ as a
function of field $F_{i}=F$, showing that $F_M \approx 3.945$ kV/cm
and $J_M \approx 3.1269$ A/cm$^2$ for $T=5$ K and the SL values of
Table~\ref{SLvalues}.}
\label{fig1}
\vspace{-0.5cm}
\end{center}
\end{figure}

For numerical treatment, it is convenient to render the equations dimensionless. We have
used the following definitions \cite{BGr05}
\begin{eqnarray}
&&\mathcal{F}_i = {F_{i}\over F_M}, \quad
 {\tilde n}_i = {n_i \over N_D}, \quad
\mathcal{J}_{i\to i+1} =  {J_{i\to i+1}\over J_M}, \quad
\mathcal{J} = {J\over J_M}, \quad
{\tilde t}= {t\over t_{0}}\equiv {J_{M}t\over\varepsilon F_M },\label{nondim}\\
&& v(\mathcal{F}_i) = { v^{(f)}(F_i)\over v_M}, \quad
\phi= {V\over V_{0}}\equiv {V\over (N+1)\, F_{M}l}, \quad
\tilde{\sigma} = \rho_{c}\sigma\equiv {F_M\,\sigma \over J_M}, \quad
v_{M} = {J_M l\over e N_{D}}.\nonumber
\end{eqnarray}
The values $F_M$ and $J_M$ are defined as the field and current density at which the
tunneling current $J_{i \to i+1}$ of (\ref{Jtuni}) reaches its first relative maximum, provided 
$n_{i}=n_{i+1}= N_{D}$. With these definitions, the model equations are
\begin{eqnarray}
&& {d\mathcal{F}_i\over d\tilde{t}} + \mathcal{J}_{i\to i+1} = \mathcal{J},
\quad i=0,\dots,N, \qquad \qquad \quad \label{eqAdim}\\
&&\tilde{n}_i = {\mathcal{F}_i - \mathcal{F}_{i-1} \over \nu} + 1,
\quad i=1,\dots,N, \qquad \qquad \quad\label{poissonAdim}
\\
&&\sum_{i=0}^N \mathcal{F}_i = (N+1)\, \phi(t), \quad t \ge 0,\qquad \qquad \quad
\label{biasAdim}\\
&&\mathcal{J}_{i\to i+1} = v(\mathcal{F}_i) \left\{\tilde{n}_i - \rho_0 \ln 
\left[1 + e^{-a \mathcal{F}_i} \left(e^{{\tilde{n}_{i+1} \over \rho_0} } - 1 
\right) \right] \right\},i=1,\dots,N-1,\label{JtuniAdim}\\
&&\mathcal{J}_{0\to 1} = \tilde{\sigma} \mathcal{F}_0, \qquad \qquad \quad
\label{J0Adim}\\
 && \mathcal{J}_{N\to N+1} = \tilde{\sigma}\mathcal{F}_N\tilde{n}_N,
 \qquad \qquad \quad\label{JNAdim}
\end{eqnarray}
where
\begin{eqnarray}
\nu = {e N_D \over \varepsilon F_M}, \quad
\rho_0 = {m^* k_B T \over \pi \hbar^2 N_D}, \quad
a = {elF_M \over k_BT},
\end{eqnarray}
are dimensionless parameters. Their values for the SL described in Table~\ref{SLvalues} 
are given in Table~\ref{typicalvalues}. Figure \ref{fig1} shows the tunneling current
density as a function of a homogeneous field profile $F_{i}=F$ when the electron densities 
are set equal to the doping density in all the SL wells. We observe several relative maxima,
the first of which yields the values of $F_{M}$ and $J_{M}$. The field intervals between two 
consecutive maxima roughly correspond to plateaus in the I--V characteristics.
\begin{table}[ht]
\caption{Typical scales for $T=5$ K.}
\begin{center} \footnotesize
\begin{tabular}{ccccccccc}
 \hline
$F_M$ & $J_M$ & $v_M$ & $x_0$ & $t_0$ & $\nu$ & $\rho_0$ & $\rho_c$ & $V_0$ \\
(kV/cm) & (A/cm$^2$) & (m/s) & (nm) & (ns) & (--) & (--) & ($\Omega$ m) & (V)\\
& & & & & & & & \\
--&-- & $\displaystyle{J_M l \over e\, N_D}$
    & $\displaystyle{\varepsilon F_M l \over e N_D}$
    & $\displaystyle{\varepsilon F_M \over J_M}$
    & $\displaystyle{e N_D \over \varepsilon F_M}$
    & $\displaystyle{m^* k_B T \over \pi \hbar^2 N_D}$
    & $\displaystyle{l F_M \over e v_M N_D}$
    & $F_M N l$\\
& & & & & & & & \\
3.945 & 3.127 & 1.691 & 2.494 & 2.066
 & 5.212 & 0.111 & 12.62 & 0.205\\
 \hline
\end{tabular}
\end{center}
\label{typicalvalues}
\end{table}

In order to numerically solve the discrete model, we first sum (\ref{eqAdim}) from $i=0$
to $N$ and use the bias condition (\ref{biasAdim}) to calculate $\mathcal{J}$. The result
is
\begin{eqnarray}
\mathcal{J} = {d \phi(t) \over d\tilde{t}} + {1 \over N+1} \sum_{i=0}^N 
\mathcal{J}_{i \to i+1}. \label{nj}
\end{eqnarray}
To solve (\ref{eqAdim})-(\ref{JNAdim}) together with the initial condition
\begin{eqnarray}
\mathcal{F}_{i}(0) = \mathcal{F}_{i0},\quad i=0,\ldots N, \quad
\phi(0)= \sum_{i=0}^N{\mathcal{F}_{i0}\over N+1}, \label{ini}
\end{eqnarray}
is equivalent to solving (\ref{poissonAdim}) - (\ref{JNAdim}) plus the following
equation instead of (\ref{eqAdim}): 
\begin{eqnarray}
{d\mathcal{F}_i \over d\tilde{t}} = {d \phi\over d\tilde{t}}
 + {1 \over N+1} \sum_{j=0,j\ne i}^N \mathcal{J}_{j \to j+1}
 - {N \over N+1} \mathcal{J}_{i \to i+1} , \quad i=0,\dots,N.\label{eqSum}
\end{eqnarray}
The new system of equations also satisfies the bias condition (\ref{biasAdim}). This can
be checked by adding all equations (\ref{eqSum}) from $i=0$ to $N$ which implies that 
$\sum_{i=0}^N \mathcal{F}_i - (N+1) \phi$ is a constant, equal to zero because of the 
initial conditions. To solve our dimensionless equations, we have used an embedded 
Runge-Kutta method of order 7(8) with step-size control and error estimate, checking the
results independently by means of an implicit BDF (backward differentiation formula) 
method of order 1 to 4, solved by means of Newton-Raphson iterations. These methods are
more accurate than those used by Amann et al \cite{AWB01}.

\section{$I$--$V$ characteristics}
\label{sec:4}
\begin{figure}[ht]
\begin{center}
\epsfxsize=160mm
\epsfbox{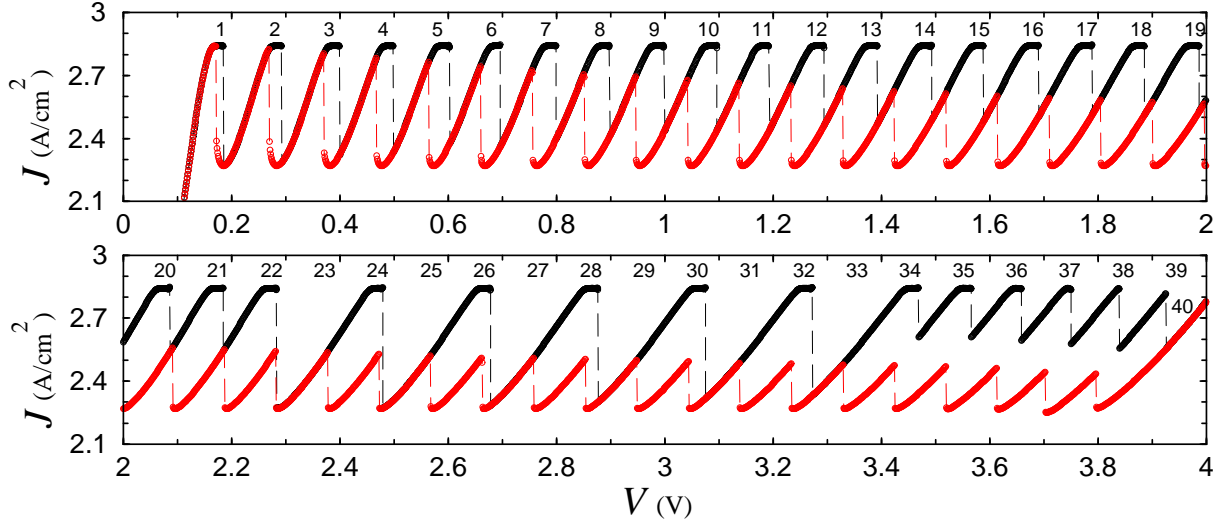}
\vspace{0.2cm}
\caption{(Color online) I--V characteristics of the 40-period 9-4 SL of Ref.~\cite{KHP97}
obtained by up- and down-sweeping adiabatic processes for $V \in [0,4]$ V. Parameters
correspond to Table~\ref{SLvalues} at 5K and with cathode resistivity of 25.2 $\Omega$m
($\tilde{\sigma}=0.5$). The branch number increases with voltage: the $i$th branch has a
CAL separating low and high field domains which is located at the $(N-i+1)$th well.}
\label{fig2}
\end{center}
\end{figure}

\begin{figure}[ht]
\begin{center}
\epsfxsize=100mm
\epsfbox{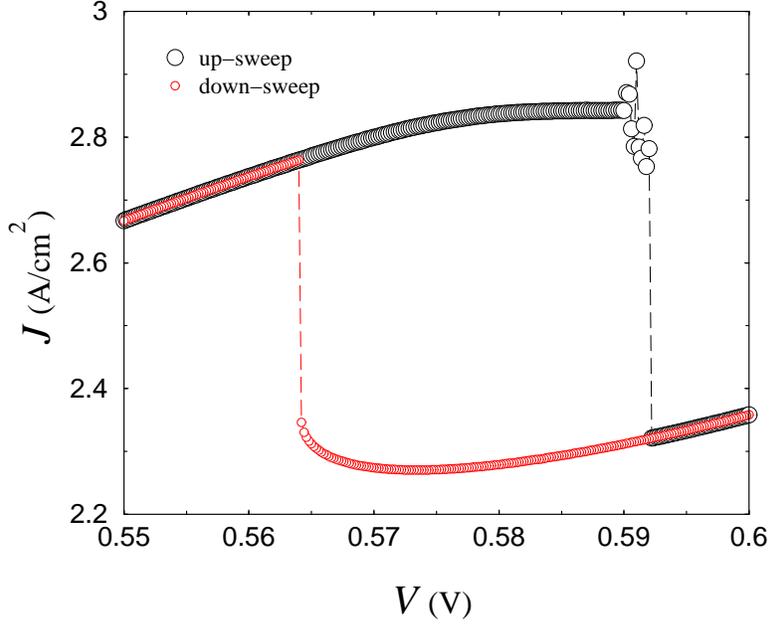}
\vspace{0.2cm}
\caption{(Color online) Bistability between Branches 5 and 6 of the I--V characteristics in 
Fig.~\ref{fig2} ($V \in [0.55,0.6]$ V). The stationary solution becomes unstable to a
small-amplitude oscillatory solution at the upper end of Branch 5.}
\label{fig3}
\end{center}
\end{figure}

We have constructed numerically the first plateau of the I--V characteristics of the SL whose 
parameters are compiled in Table \ref{SLvalues} at a temperature of 5K and a contact 
resistivity $1/\sigma=25.2$ $\Omega$m ($\tilde{\sigma}=0.5$). We depict in Fig.~\ref{fig2}
the portion of the stable static branches obtained by voltage up-sweep from zero
volts to the end of the first plateau and also the portion obtained by voltage down-sweep 
from high voltage values. Note that the upper parts of odd numbered branches (from the 23rd
to the 33rd) do not appear because they correspond to regions of bistability and are skipped 
during voltage up-sweep. Likewise, the central parts of branches 35th to 39th are not shown
because they correspond to regions of tristability and are skipped during voltage up-sweep
and down-sweep. All these curves can be entirely shown by up-sweeping starting from the
parts thereof shown in the figure. The voltage distribution for different locations of the 
domain boundary for a fixed applied voltage in regions of multistability is shown in 
Figures 8 and 9 of the review paper \cite{BGr05}, which contains further background
discussions and references to pertinent literature. The basin of attraction of the static branches 
shown in Fig.~\ref{fig2} is very small near their ends and it is mainly in these regions that 
our high precision numerical methods make the difference with previous calculations; see 
Fig.~\ref{fig3}. We have found that the I--V branches may undergo Hopf bifurcations
to small-amplitude oscillatory solutions near their upper ends, as we explain in Section
\ref{sec:6}.

According to their stability feature, we can distinguish four different types of I--V branch:
\begin{enumerate}
\item The first branch $B_1$ is singly stable from $V=0$ until its upper part coincides with 
the lower end of branch $B_2$, giving rise to a narrow region of bistability.
\item From branch $B_2$ to $B_{21}$, we observe that the branch length increases (the 
extension of $B_2$ is 0.122 V whereas that of $B_{19}$ is 0.177 V) while their slope (the 
PDC) decreases. We see that the low voltage branches have a steep slope similar to that below 
the first maximum of the homogeneous tunneling current density in Fig.~\ref{fig1}, whereas
the high voltage branches have a smaller slope similar to that of the low part of the ascent
to the second maximum in Fig.~\ref{fig1}. Why? The field profile in the low voltage 
branches is similar to that of a solution with spatially uniform field, which therefore obeys
$J_{i\to i+1}(F,N_{D},N_{D}) = J$, thereby corresponding to the first branch of the curve 
in Fig.~\ref{fig1}. Similarly, high voltage branches are close to spatially homogeneous 
field profiles satisfying the same relation, but now the field profile corresponds to the third
branch of Fig.~\ref{fig1}. Branches with intermediate voltages are a combination of low 
and high field domains, and therefore their slopes are interpolations between low and high
PDC. The central part of each branch is singly stable, 
while their two ends are bistable. As the branch number increases, the central part shrinks 
and the bistable regions including the ends of the branch grow.
\item Branch $B_{22}$ is the last one having a central singly stable region and the first one
with a tristability region near its upper end. In the tristability region, branch $B_{22}$ 
coexists with the central part of branch $B_{23}$ and the lower part of branch $B_{24}$.

\item Branch $B_{23}$ is bistable except in its central part and its upper end where it is
tristable. The lower part of this branch is still bistable.

\item Branches $B_{24}$ to $B_{39}$ are tristable in their central parts and ends, but they
have two bistability regions. As voltage increases, the tristability regions grow at the expense
of the bistability regions. 

\item The last branch of the first plateau, $B_{40}$, has a tristable lower end, a bistable 
central region and it is singly stable from there until it reaches the second plateau in its
upper end.
\end{enumerate}

\section{Large switching: evolution of $J(V)$ along the $I$--$V$ curve}
\label{sec:5}
In this Section, we describe the dynamical response of the SL to a voltage switching $V(t)= 
V_{i} + \Delta V\, t\, H(\tau_{r}-t)/\tau_{r}+ \Delta V\, H(t-\tau_{r})$, in which 
$\Delta V$ is constant, and $H(t)=1$ if $t>0$, $H(t)=0$ if $t<0$ is the Heaviside unit step 
function. We select the initial voltage in 
the central part of one branch and the final voltage $V_{f}=V_{i}+\Delta V$ in different 
parts of another stable branch, so that several branches can be found between $V_{i}$
and $V_{f}$. We have found different scenarios.

\subsection{Switching from bistable branches: modified tripole-dipole scenario}
Let us choose $V_{i}= 0.83$V, in the central singly stable part of branch $B_{8}$ and
$V_{f}=1.37$V, on the central part of branch $B_{14}$. These branches are bistable but they 
have a central part for whose voltages no other static solution is stable (the branch is singly 
stable there). Fig.~\ref{fig4} shows the dynamical response of the total current density $J(t)$
to voltage switching with two different ramping times. For large enough ramping times (not shown),
the current density $J(t)$ follows adiabatically the I--V characteristics. Below a first critical
value of the ramping time, $J(t)$ cannot reach the upper end of the branches and it falls to the
lower part of the following branch in the voltage range where both branches are stable.
{\em For each branch crossing}, this fall occurs via the following modified tripole-dipole
scenario whose current trace is depicted in Fig.~\ref{fig5}:
\begin{figure}[ht]
\vspace{0.1cm}
\centerline{\hbox{
       \epsfxsize=80mm
       \epsfbox{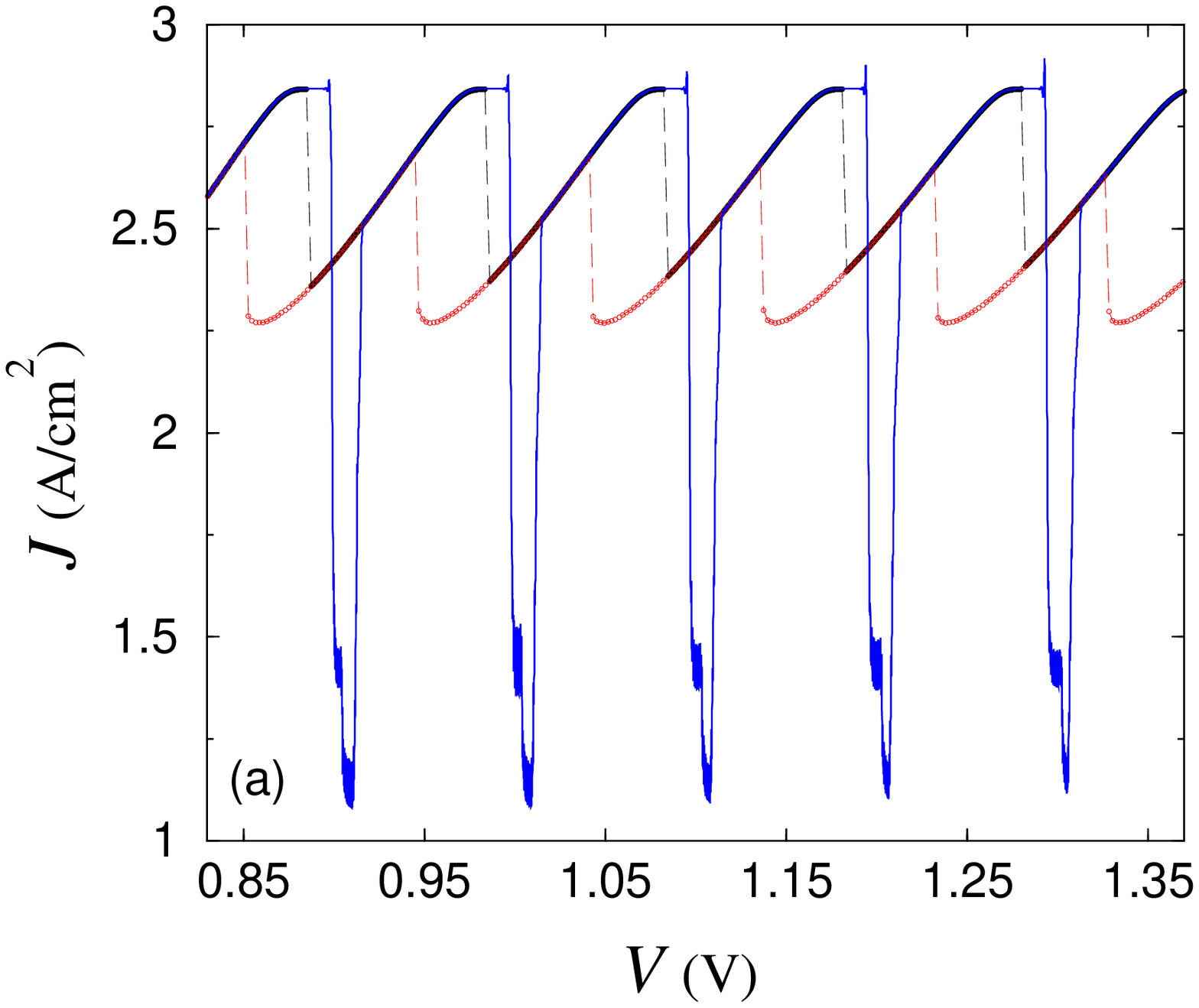} \hspace{0.8cm} 
       \epsfxsize=80mm
       \epsfbox{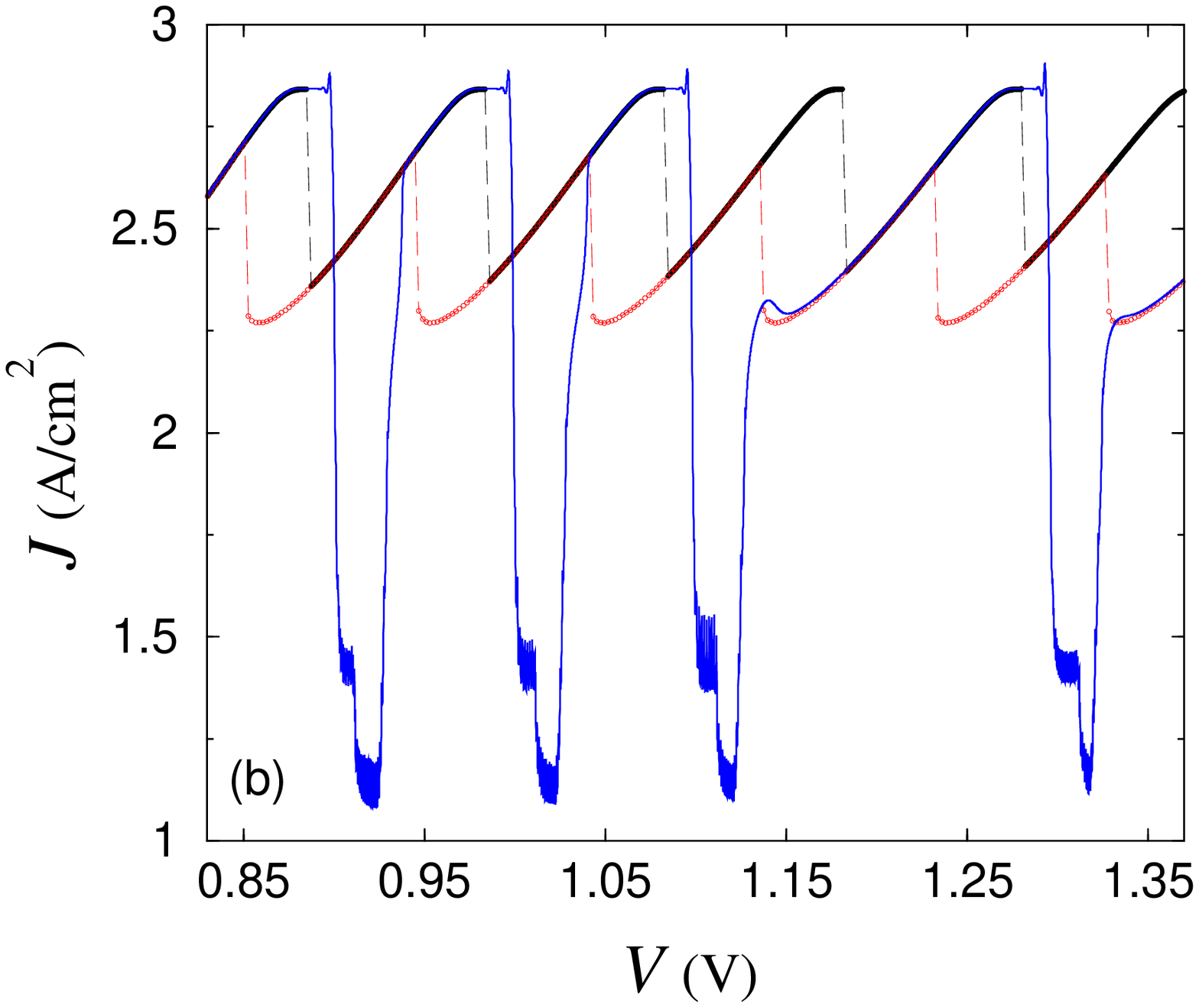}}}
\caption{(Color online) Total current density during voltage switching from $V_{i}=0.83$V
to $V_{f}= 1.37$V (seven branches) for ramping times: (a) $\tau_{r} = 30 \mu$s (dimensionless value:
$\tilde{\tau }_{r} = 2 \times 10^4$), and (b) $\tau_{r} = 15 \mu s$ ($\tilde{\tau}_{r} = 10^4$).
Thick black line: upper part of the I--V branches, red thin line: lower part of the 
static I--V branches, thin blue line: response curve $(V(t),J(t))$.
In (a), the current follows adiabatically the I--V curve, whereas in (b), the ramping time
is so short that the 4$^{th}$ and 6$^{th}$ branches are skipped.
The dimensionless cathode conductivity is $\tilde{\sigma} = 0.5$.}  
\label{fig4}
\end{figure}

\begin{figure}[ht]
\begin{center}
\epsfxsize=100mm
\epsfbox{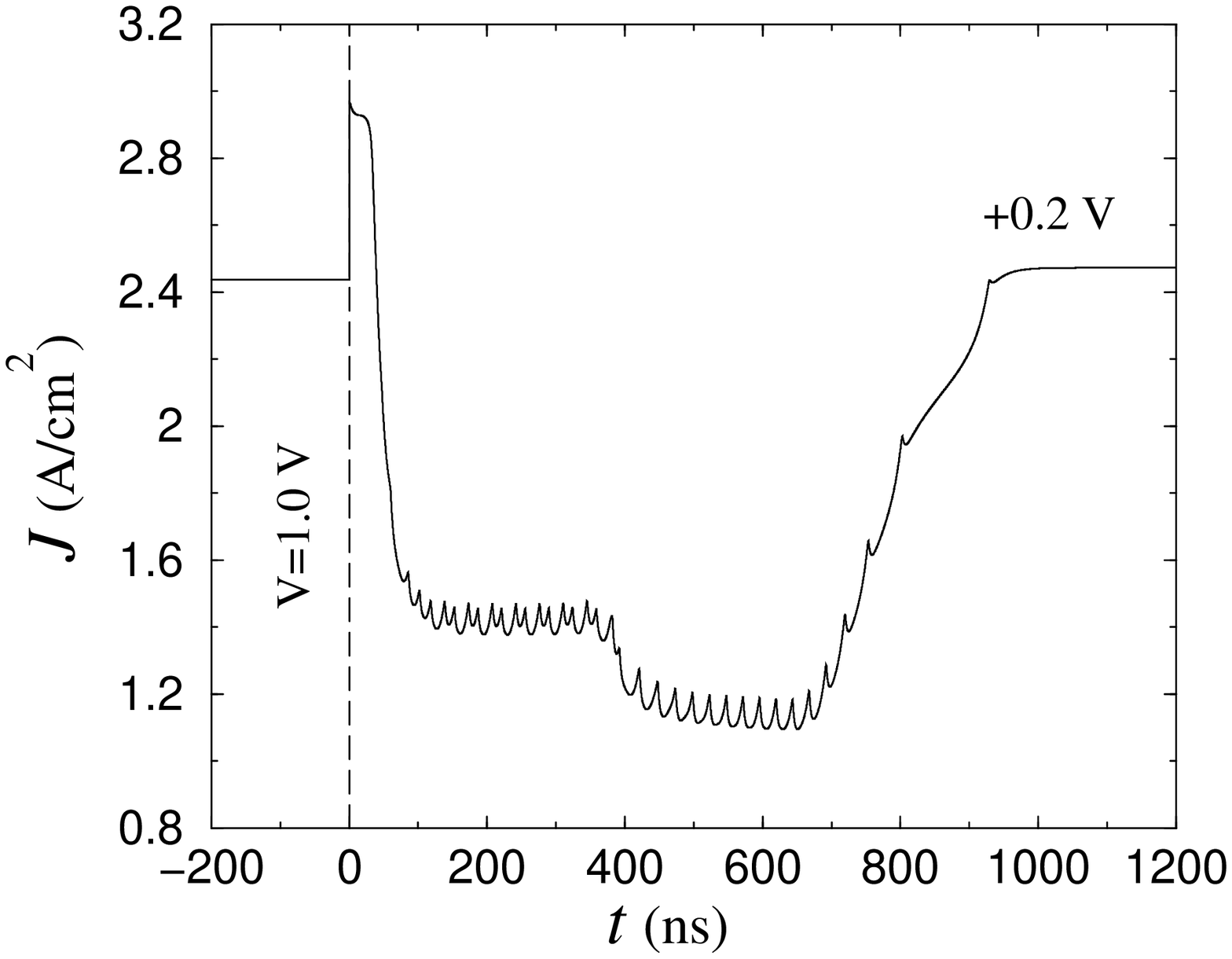}
\vspace{0.2cm}
\caption{Time trace of the total current density $J(t)$ after sudden voltage switching 
characterizing the tripole-dipole scenario. Here $V_{i}= 1.0$ V (for $t < 0$)  and
$\Delta V=0.2$ V.}
\label{fig5}
\end{center}
\end{figure}

\begin{enumerate}
\item[Phase 1] The CAL separating low and high field domains stay in the same well while
the current density increases until it surpasses the critical value at which $J=\sigma F$
intersects the curve in Fig.~\ref{fig1}, $J=J_{i\to i+1}(F,N_{D},N_{D})$ on the second 
branch thereof ($J_{c}\approx 2.9$ A/cm$^2$).
\item[Phase 2] A charge dipole wave is created at the cathode and it starts moving to the
anode while the old CAL also moves towards the anode. After a short transient during which
the current decreases, the whole structure, a {\em charge tripole} moves rigidly towards the 
anode.
\item[Phase 3] The old CAL reaches the anode leaving a charge dipole moving (after a short 
transient during which the current further decreases) rigidly towards the anode at a lower 
speed than the tripole.
\item[Phase 4] The front part of the dipole, a CDL, reaches the anode leaving only the CAL 
in its back part. The current density increases to higher values corresponding to the next
stable static solution while the CAL moves towards its final destination separating low and 
high field domains of the stable static field profile.
\end{enumerate}
Phase 2 is characterized by double-peaked current spikes (corresponding to the well-to-well
jumps of the two CAL) while Phases 3 and 4 exhibit single current spikes; see 
Fig.~\ref{fig5}. Phases 2, 3 and 4 are exactly as described in \cite{AWB01} (who did
not describe Phase 4) and corrected in \cite{RTG02} (who added Phase 4). In these 
previous works, Phase 1 was characterized by a one-well motion of the CAL towards the 
cathode. Note that experimental observations refer only to the behavior of the current density,
not to the motion of domain walls, and therefore they cannot discriminate between our Phase 1
and the different one reported in \cite{AWB01}. To explain the differences in Phase 1, we 
note that, according to Fig.~3 of Ref.~\cite{CBW00}, a CAL may move against the 
electron flow if the current is large enough and the doping density is sufficiently large. 
However, the critical current for this motion would be exponentially close to the maximum 
current for which CAL exist if the doping density is even larger, thereby eliminating in 
practice the possibility for CAL to move against the electron flow under current bias. In the 
simulations by Amann et al, the interval of currents allowing CAL motion against the electron 
flow was relatively wide (cf.\ their Figure 4), whereas we found that it was negligible for 
the tunneling current and doping density used in the present calculations. Then the increase 
of voltage due to switching is compensated by simply a current increase in Phase 1, without 
CAL motion, as indicated in Fig.~\ref{fig5}. When the current surpasses its critical value 
(see the current spike over the maximum current value for static branches in Fig.~\ref{fig4}),
Phase 2 begins: a dipole is shed from the cathode and it moves on towards the anode together 
with the old CAL. This and subsequent Phases are as in the previous works 
\cite{AWB01,RTG01}. The behavior of the current can be explained using singular 
perturbation ideas as described in the reviews \cite{Bon02,BGr05}. Since the final position 
of the CAL approaches the cathode as the voltage increases, the dipole trip becomes shorter 
as shown in Fig.~\ref{fig4}.

\begin{figure}[ht]
\vspace{0.1cm}
\centerline{\hbox{
       \epsfxsize=85mm
       \epsfbox{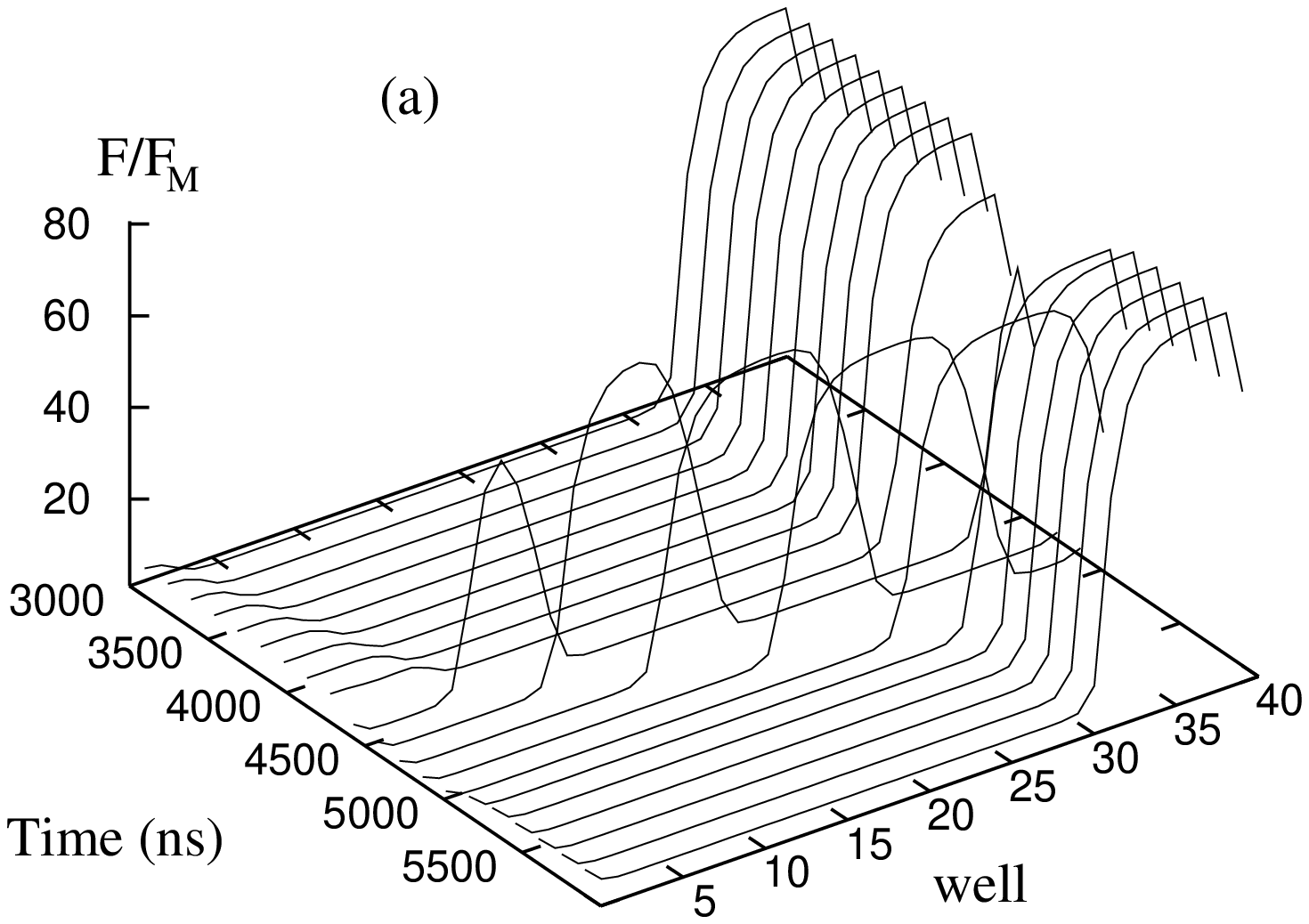} 
       \epsfxsize=85mm
       \epsfbox{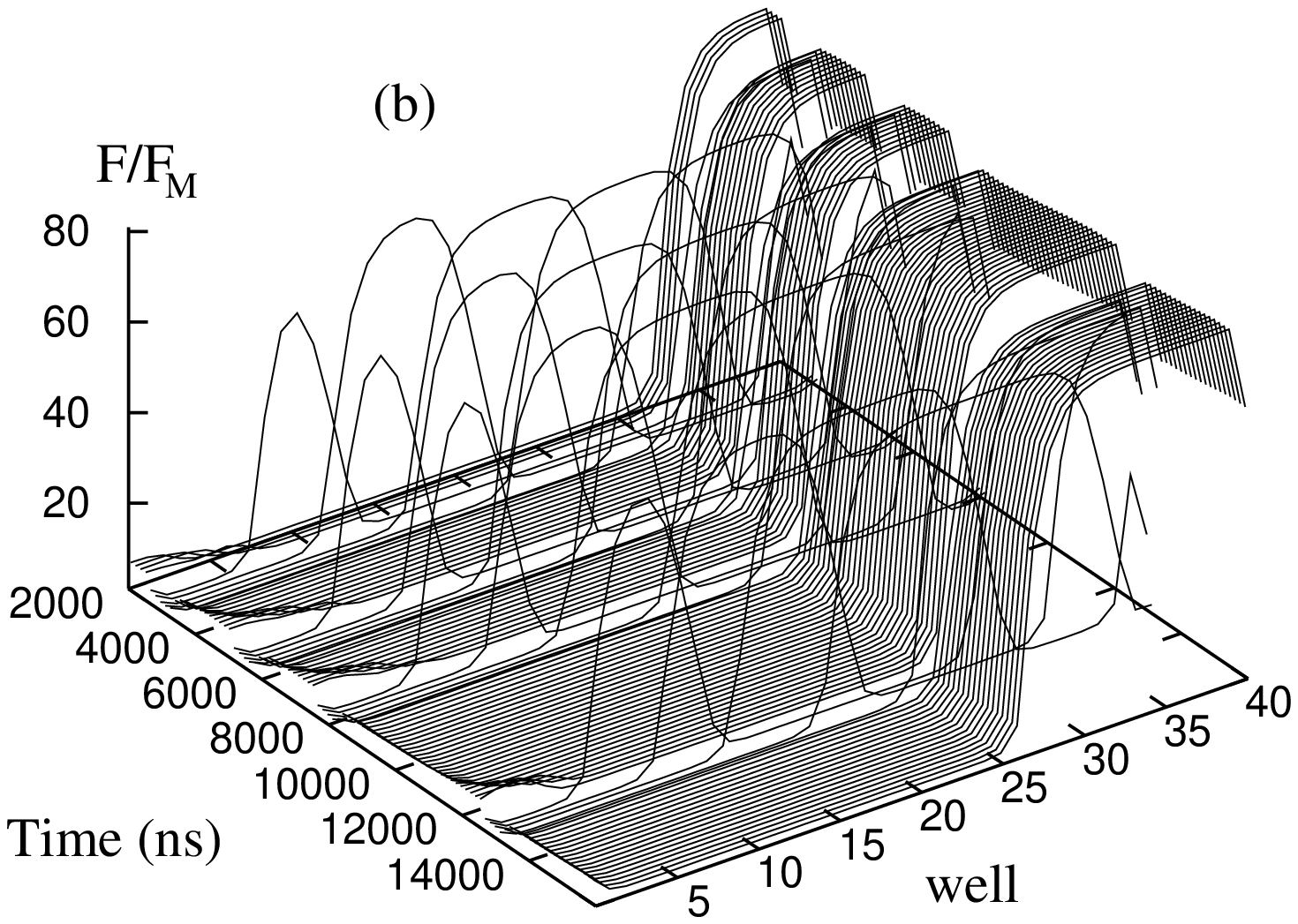}}}
\caption{Field profiles corresponding to Fig.~\ref{fig4} during relocation of the domain 
wall. (a) Detail of the emission and the travel of the CAL during the second relocation stage in
Fig.~\ref{fig4}(a). (b) The 4$^{th}$ branch in Fig.~\ref{fig4}(b) has been skipped,
therefore the 4$^{th}$ CAL is missing and there is no relocation stage.}
\label{fig6}
\end{figure}

We have seen that switching from $B_{8}$ to $B_{14}$ occurs by a succession of 
tripole-dipole scenarios. What happens if we continue decreasing the ramping time? It turns 
out that we start skipping branches. Note that the tripole-dipole process is quite fast in
Fig.~\ref{fig4}(a) and it ends at a voltage smaller than the bistability interval
near the end of the corresponding I--V branch. At that voltage, only one branch is stable and
$J(t)$ follows this static branch until the next dipole emission. However, recall that the 
bistability intervals grow at the expense of the singly stable central part of the branches as 
their voltage increases. When the ramping time is decreased below a critical value (which
depends on the branch), the voltage at the end of a process of dipole emission and travel may 
be in the bistability range of two branches. The CAL of the dipole then stops at the position
corresponding to the static branch with lower current which is closer to the cathode than the
CAL of the branch with higher current. This is observed in Fig.~\ref{fig4}(b). Note that the
branches $B_{11}$ and $B_{13}$ have been skipped. Fig.~\ref{fig6}(b) shows that the
corresponding tripole-dipole process is longer when one branch has been skipped.

\subsection{Same switching range for bistable and tristable branches}
\begin{figure}[ht]
\vspace{0.1cm}
\centerline{\hbox{
     \epsfxsize=80mm
       \epsfbox{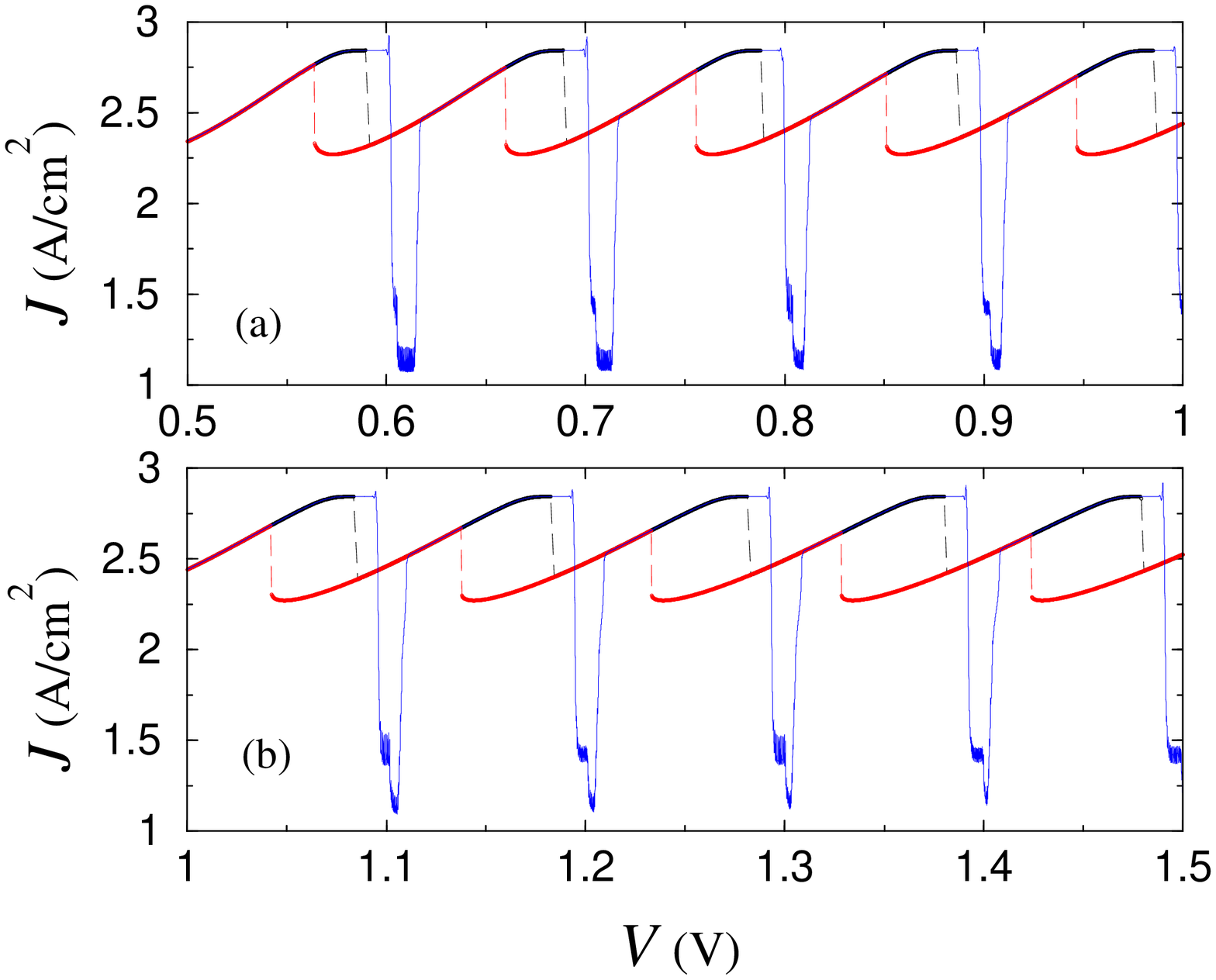} \hspace{0.8cm}
      \epsfxsize=80mm
       \epsfbox{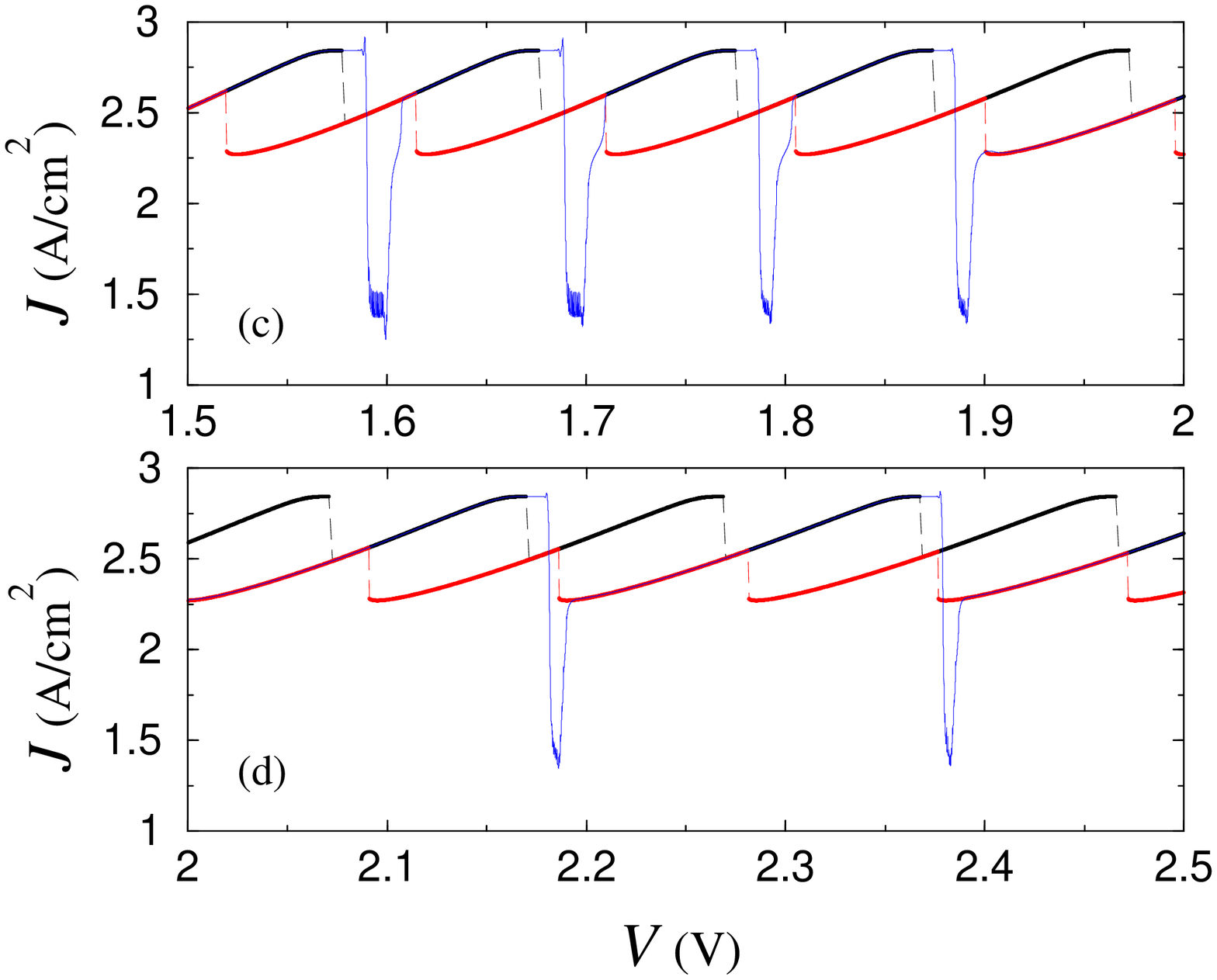}}}
\centerline{\hbox{
       \epsfxsize=80mm
       \epsfbox{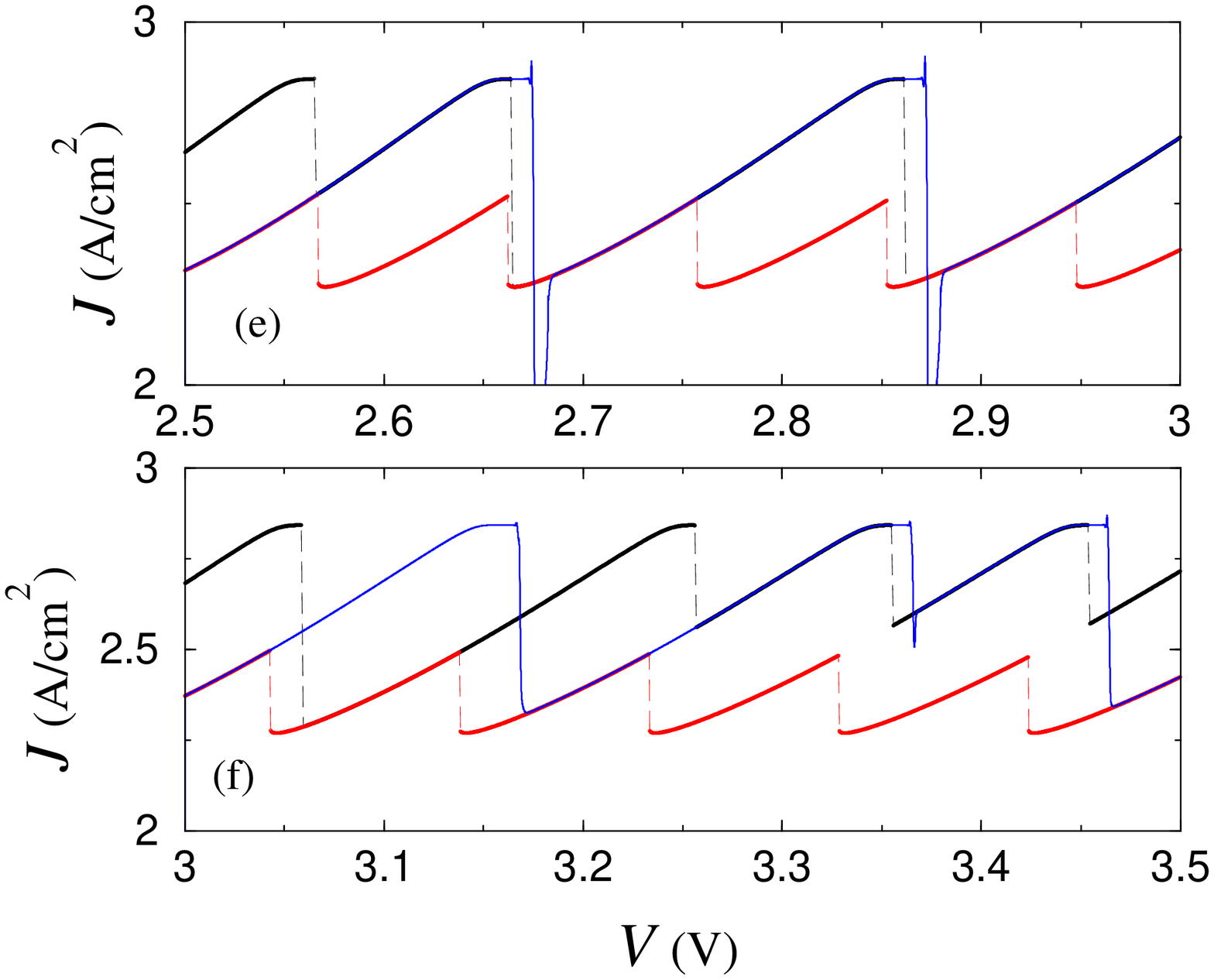} \hspace{0.8cm}
       \epsfxsize=80mm
       \epsfbox{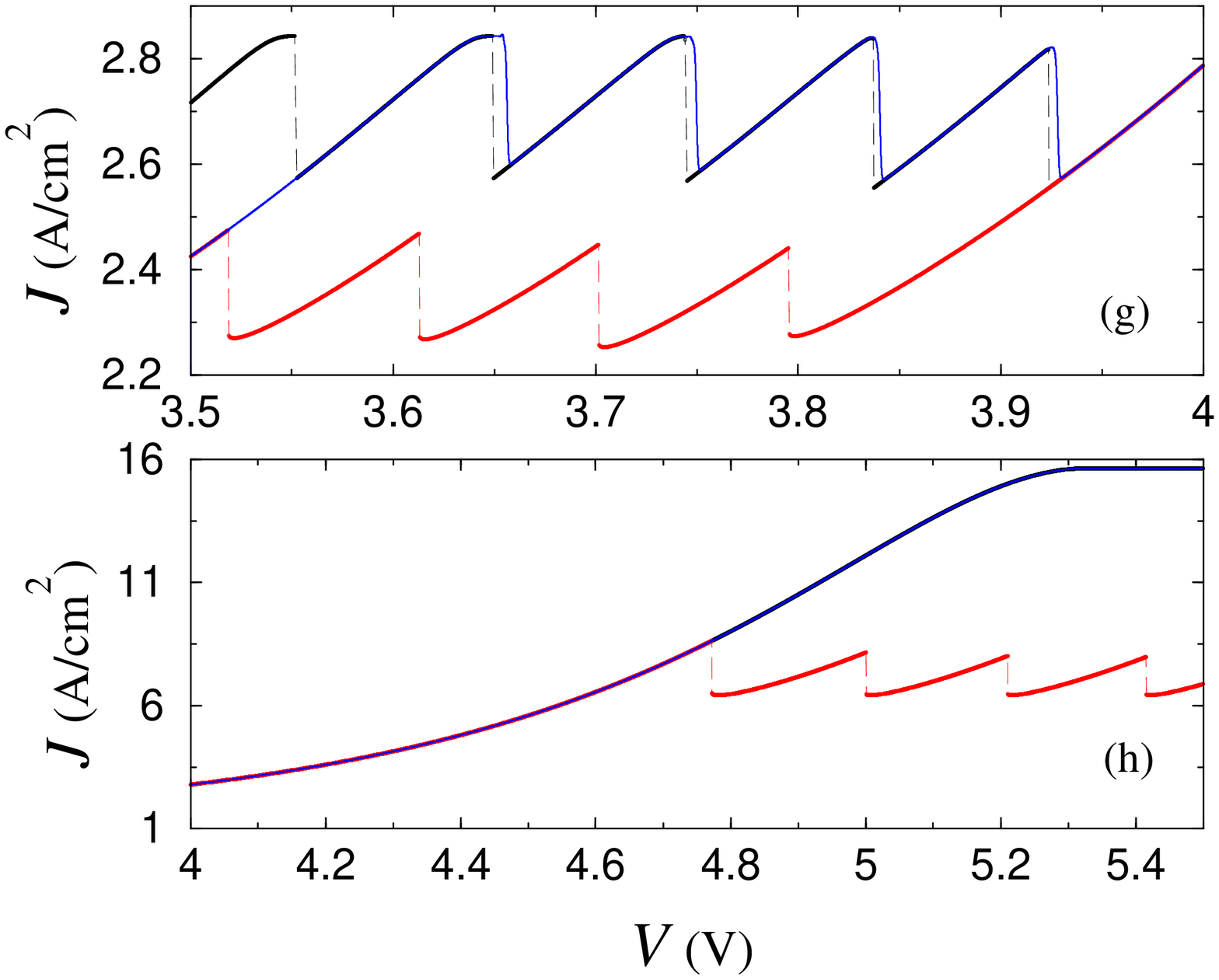}}}
\caption{(Color online) Voltage switching with $\Delta V=0.5$ V and ramping time
$\tau_{r} = 30\, \mu$s for different $V_{i}$ on the I--V characteristics:
(a) $V_i = 0.5$ V, (b) $V_i = 1$ V, (c) $V_i = 1.5$ V, (d) $V_i = 2$ V,
(e) $V_i = 2.5$ V, (f) $V_i = 3$ V, (g) $V_i = 3.5$ V, (h) $V_i = 4$ V,
$\Delta V = 1.5$ V. In  (a), (b), (g) and (h), the current follows the I--V curve along the 
upper part of each branch, whereas $J$ ranges from 6 to 16 A/cm$^2$.
In (c), a branch (the 5$^{th}$) is skipped for the first time.
In (d) and (e), branches 3 and 5 are skipped. In (f), the upper part of the 5$^{th}$  
branch is reached but the 6$^{th}$ is skipped.}
\label{fig7}
\end{figure}

It should be clear from the previous discussion that the intervals of bistability have great 
influence on the dynamic response to voltage switching. To make this clearer, we have 
depicted in Figures~\ref{fig7}(a) to (h) the dynamic response to a voltage switching of width
$\Delta V=0.5$ V with ramping time $\tau_{r}= 30 \mu$s (as in Fig.~\ref{fig4}(a)) 
for different initial voltages $V_{i}$. We observe that the tripole-dipole scenario occurs for 
small $V_{i}$, branches start to be skipped as $V_{i}$ increases, and the tripole-dipole 
process disappears at even larger voltages at which the tristability range of the branches
is very large. Of course, the occurrence of the tripole-dipole scenario depends on the ramping
time: it reappears again as the ramping time decreases sufficiently.

\section{Voltage switching to $V_{f}$ near the end of a branch}
\label{sec:6}
In this Section, we report the current response to voltage switching from $V_{i}$ to a 
voltage $V_{f}$ close to the end of the same branch. Firstly, we shall select branches near the 
end of the plateau, which were the only ones considered in Ref.~\cite{AWB01} (see
Section VI). Secondly, we shall select branches near the beginning of the plateau and observe
a rather different behavior.

\subsection{High voltages near the end of the I--V plateau} 
\begin{figure}[ht]
\begin{center}
\epsfxsize=100mm
\epsfbox{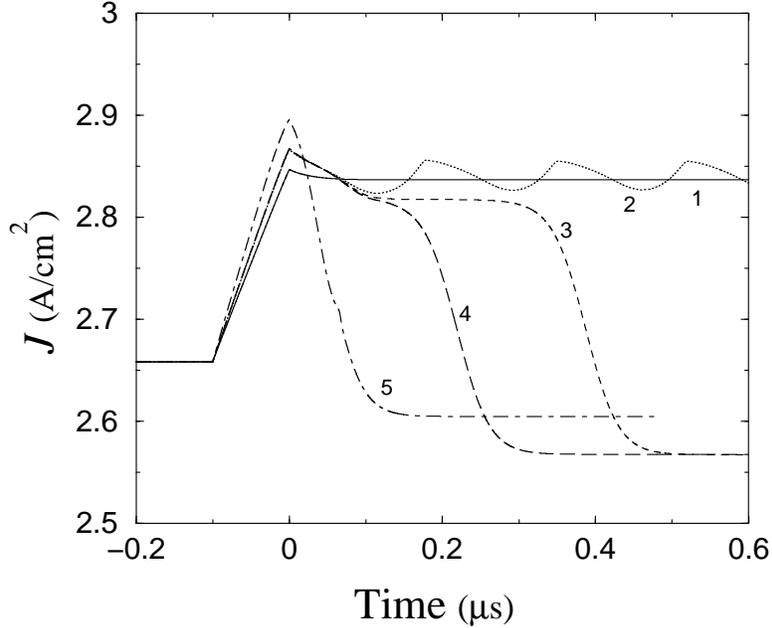}
\caption{Current response to voltage switching from $V_{\rm i}=3.68$ V to five 
different values of $V_f$ near $V_{\rm th}$ (1: 3.74, 2: 3.747, 3: 3.7472732, 4: 3.7474 and
5: 3.76 V). The ramping time is $\tau_r=100$ ns and we have set $t=0$ at the end of voltage 
switching. Note the oscillatory behavior for $V_f = 3.747$ V.}
\label{fig8}
\end{center}
\end{figure}

\begin{figure}[ht]
\begin{center}
\epsfxsize=100mm
\epsfbox{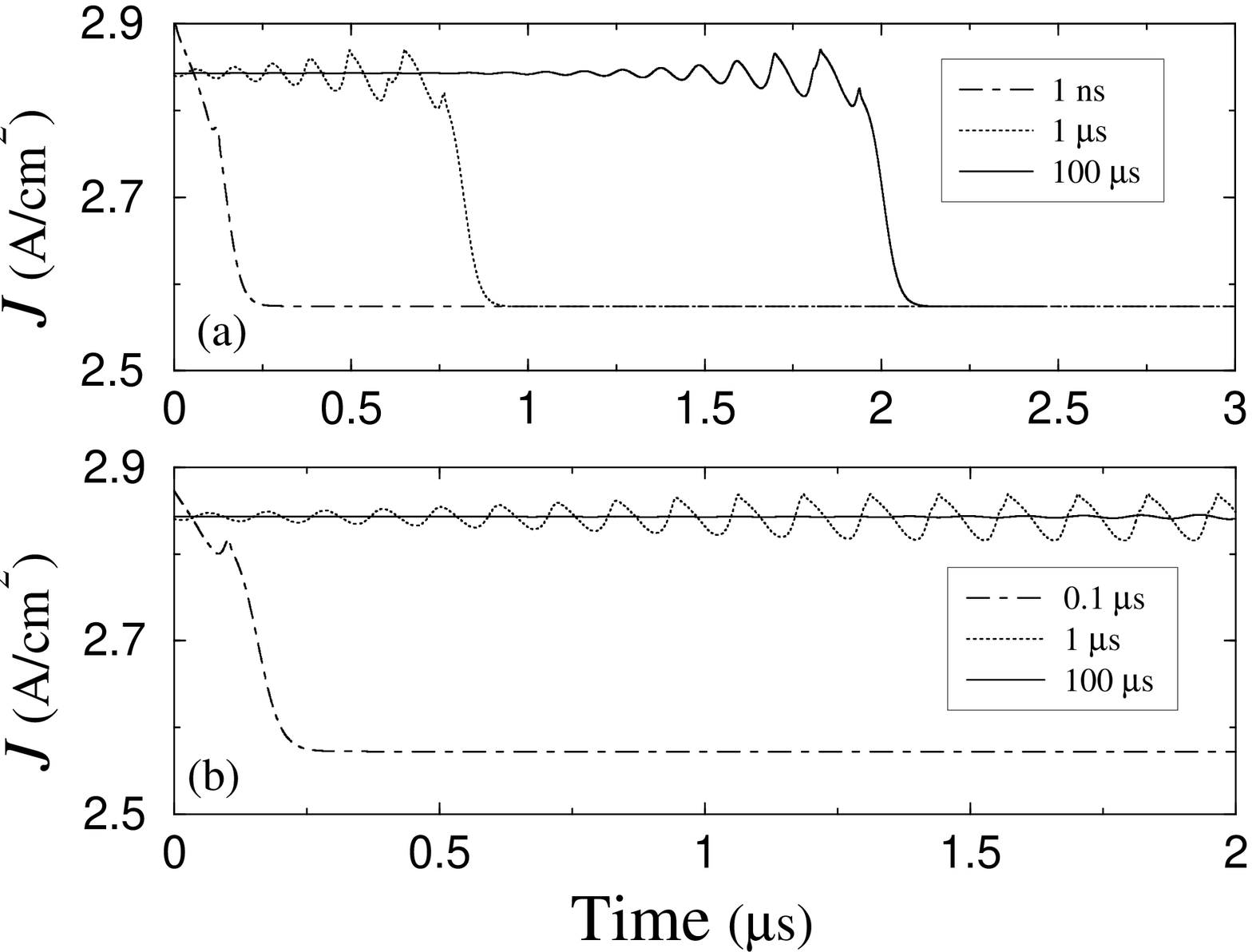}
\caption{Current response to voltage switching for $V_{\rm i}=3.5$ V, two different
values of $V_f$ and three different ramping times in each case:
(a) $V_f=3.555$ V $>V_{\rm th}$, for $\tau_r = 1$ ns, 1 $\mu$s and 100 $\mu$s.
(b) $V_f=3.5545$ V $<V_{\rm th}$, for $\tau_r = 0.1$ $\mu$s, 1 $\mu$s and 100 $\mu$s.
In all cases, we set $t=0$ at the end of voltage switching. For $\tau_r = 100$ $\mu$s, the
current undergoes a small-amplitude oscillation in the case of Fig.~\ref{fig9}(b).}
\label{fig9}
\end{center}
\end{figure}

Fig.~\ref{fig8} depicts the current response to a voltage switching starting with a 
$V_{i}$ in the bistable part of branch $B_{37}$, closer to its end. Keeping a ramping time 
of 100 ns, we observe similar behavior to that reported by Amann et al \cite{AWB01}: 
(i) if $V_{f}<V_{th}$ (the end of the static branch), the current remains on the same branch 
but the time it takes to settle in its final value increases as $V_{f}$ approaches $V_{th}$; 
(ii) if $V_{f}>V_{th}$, the final state is on the next branch, and the transient stage lasts 
longer as we approach $V_{th}$, cf.\ Fig.~\ref{fig9}(a). Similarly, the longer the ramping 
time is, the longer the transient before the current drops to that of the following I--V branch 
seems to be, as indicated in Fig.~\ref{fig9}(b). This Figure shows the influence of the 
ramping time on the current response to voltage switching with $V_{f}$ close to $V_{th}$ 
for I--V branches near the end of the plateau. We have selected now Branch $B_{35}$
and changed the ramping time from small to large for two different $V_{f}$ close to 
$V_{th}$, one larger than $V_{th}$, the other smaller. For $V_{f}>V_{th}$, we observe that
the current eventually drops to the lower value on branch $B_{36}$, but the transient stage
lasts longer as the ramping time increases. For $V_{f}\approx V_{o}<V_{th}$, the basin of
attraction of $B_{35}$ is so small that the current eventually drops to its final value on Branch
$B_{36}$. However, the way in which this happens depends strongly on the ramping time: 
(i) if the ramping time is too small, the current drops rapidly and the final state is on Branch 
$B_{36}$ is reached soon; (ii) for intermediate ramping times, the current oscillates about the 
static Branch $B_{35}$ before it falls to Branch $B_{36}$, and (iii) for large ramping times, 
the current seems to settle down to the static value on Branch $B_{35}$ and it resists much 
longer before it eventually drops down to Branch $B_{36}$. 

\begin{figure}[ht]
\centerline{\hbox{
\epsfxsize=95mm
\epsfbox{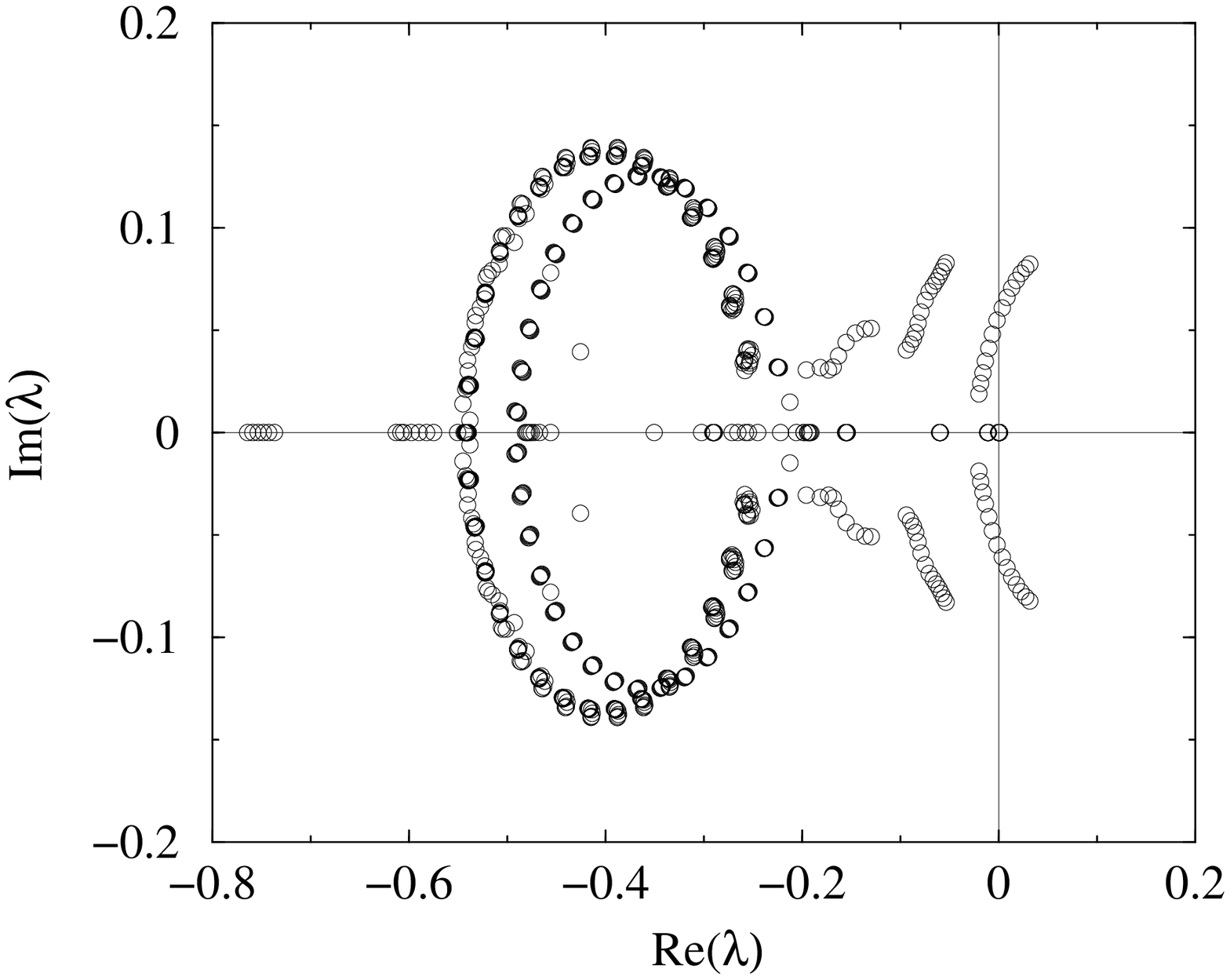}
}}
\caption{Evolution of the 41 eigenvalues determining linear stability of the static branch 
$B_{35}$ for 20 values of the applied voltage 
from $V=3.54$ to $3.56$ V.}
\label{fig10}
\end{figure}

A distinct feature of the current response to voltage switching with $V_{f}$ near the end of a 
static branch is that, in case (i), the final stable state for $V_{f}<V_{th}$ very close to 
$V_{th}$, may be oscillatory, not stationary. This is suggested by the oscillations in 
Figs.~\ref{fig8} and \ref{fig9}, and further confirmed by the linear stability analysis of 
Appendix \ref{appA}. There it is shown that the static branch loses stability at some $V_{o}
<V_{th}$ because the real part of two complex conjugate eigenvalues becomes positive for 
$V_{o}<V<V_{th}$. This is clearly seen in Fig.~\ref{fig10}, which depicts the eigenvalues 
of the linear stability problem about the static branch $B_{35}$ for 20 different voltage 
values close to $V_{th}$. For this branch, $V_{o}\approx 3.5547$ V.

\subsection{Current response to switching near the end of low voltage branches}
\begin{figure}[ht]
\begin{center}
\epsfxsize=100mm
\epsfbox{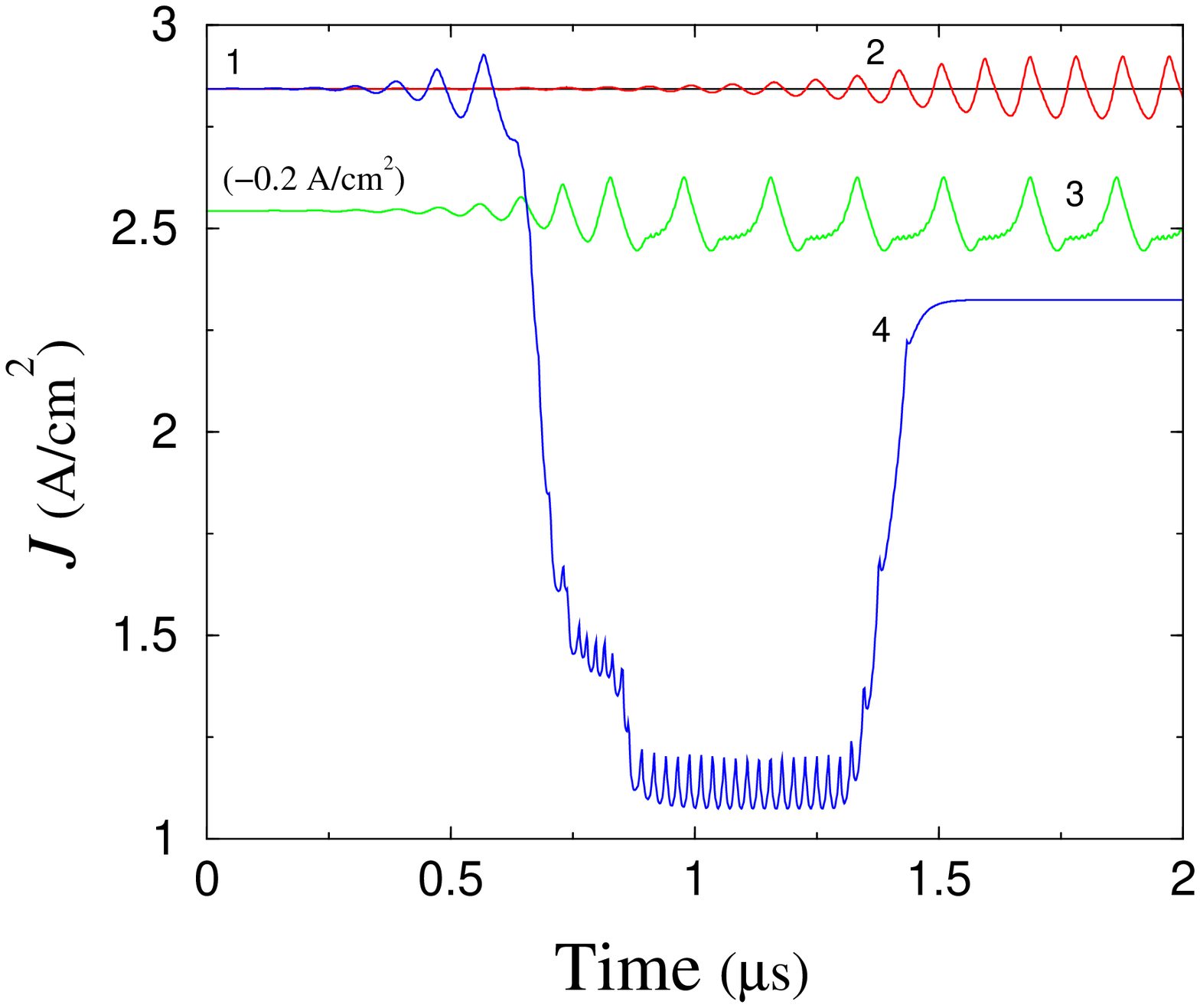}
\vspace{0.1cm}
\caption{(Color online) Current response to voltage switching between Branches 5 and 6 of 
the I--V characteristics in Fig.~\ref{fig2} for $V_i = 0.56$ V, ramping time $\tau_r = 10 \,
\mu$s, and four different values of $V_f$ near $V_{th}$: 1: 0.59, 2: 0.591, 3: 0.59203 and 4: 
0.593 V. Note that curve 3 is shifted down 0.2 A/cm$^2$ for the sake of clarity.}
\label{fig11}
\end{center}
\end{figure}

Fig.~\ref{fig11} shows that switching near the end of low voltage I--V branches is more
complex than that described previously. It turns out that the current drop to the next branch 
may occur via the tripole-dipole scenario, unlike in the numerical simulations
by Amann et al (who always selected high voltage static branches having intervals of 
tristability), but according to the experimental observations by Rogozia et al
(cf.\ Fig.~9 of Ref.~\cite{RTG02}). Another discrepancy between the numerical simulations of 
Ref.\ \cite{AWB01} and experiments is that the current spike accompanying dipole 
emission is much taller in the simulations (twice the maximum current of the static branches 
instead of the experimentally observed 20\% increase). In our simulations, the current spike 
accompanying dipole emission is much smaller than in the previous calculations by Amann et 
al, but this is due to the different cathode conductivity and critical current for dipole creation. 

\begin{figure}[ht]
\begin{center}
\epsfxsize=100mm
\epsfbox{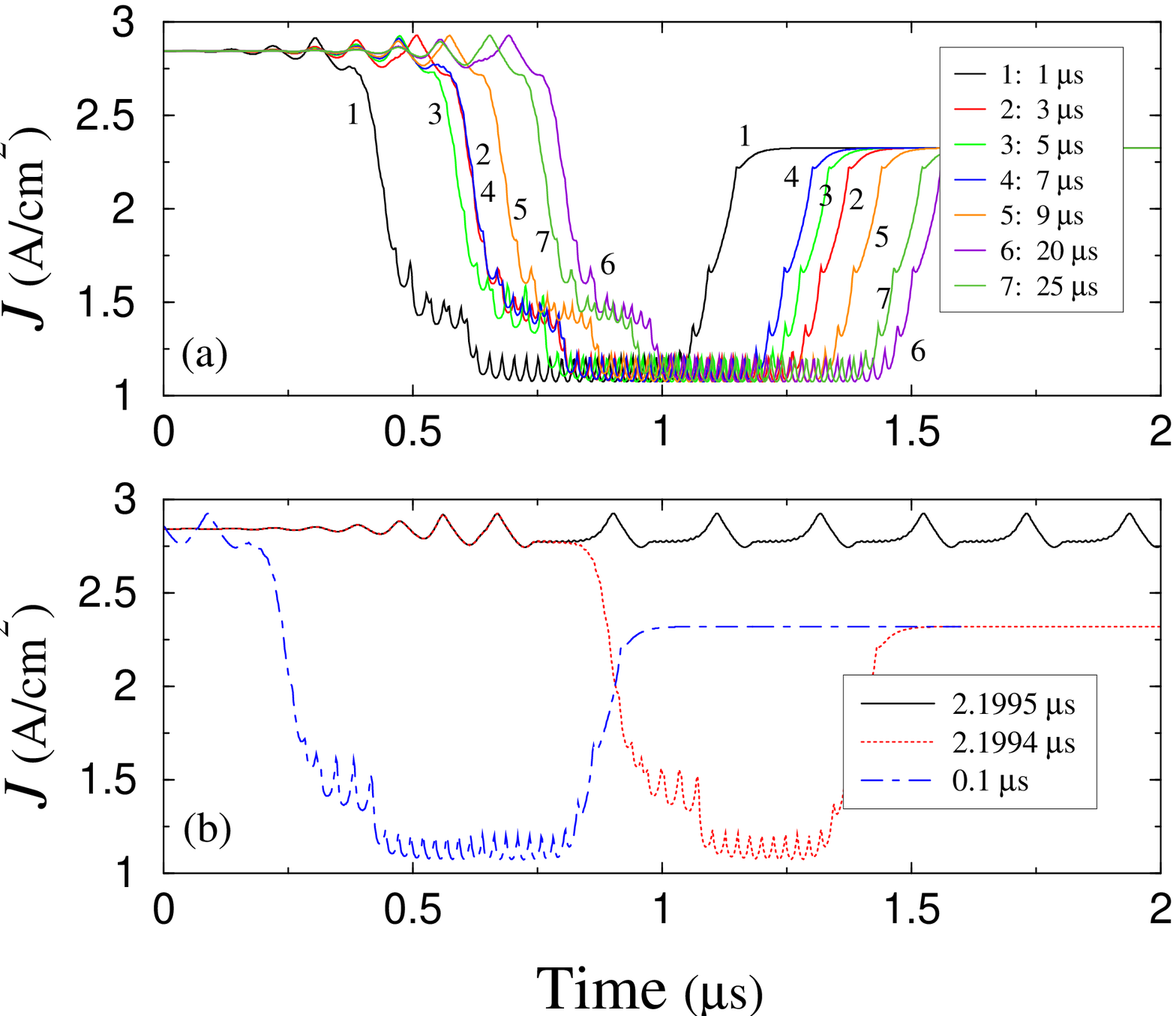}
\caption{(Color online) Current response to voltage switching between Branches 5 and 6
of the I--V  characteristics in Fig.~\ref{fig2} for $V_i = 0.56$ V.
(a) $V_f = 0.593$ V $>V_{th}$ and seven different ramping times from 1 to 25 $\mu$s.
(b) Details of current response for $V_f = 0.59203$ V $<V_{th}$ and three short ramping
times: 0.1, 2.1994 and 2.1995 $\mu$s.}
\label{fig12}
\end{center}
\end{figure}

The appearance of the tripole-dipole scenario during voltage switching occurs up to voltages 
corresponding to branches starting to display tristability. If we fix $V_{f}$ sufficiently close
to $V_{th}$, $V_{f}<V_{th}$, and change the ramping time, we observe a peculiar behavior.
For all ramping times, the relocation of the domain wall separating the low and high field 
parts of the field profile happens via the tripole-dipole scenario, as shown in 
Fig.~\ref{fig12}(a). However the time $t_{d}$ at which the tripole-dipole scenario starts is 
not a monotone function of the ramping time: it seems that $t_{d}$ may have local maxima
and minima as a function of $\tau_{r}$. We have observed that $t_{d}$ increases with
 $\tau_{r}$ up to $\tau_{r}\approx 2.2 \mu$s. Then $t_{d}$ decreases  with $\tau_{r}$ up
 to at least 7 $\mu$s. Then $t_{d}$ increases again for larger $\tau_{r}$, as indicated in 
 Fig.~\ref{fig12}(a). We have checked that $t_{d}$ is again smaller for a ramping time of 
 100 $\mu$s than for $\tau_{r}= 30 \mu$s. 

\begin{figure}[ht]
\begin{center}
\centerline{\hbox{
\epsfxsize=90mm
\epsfbox{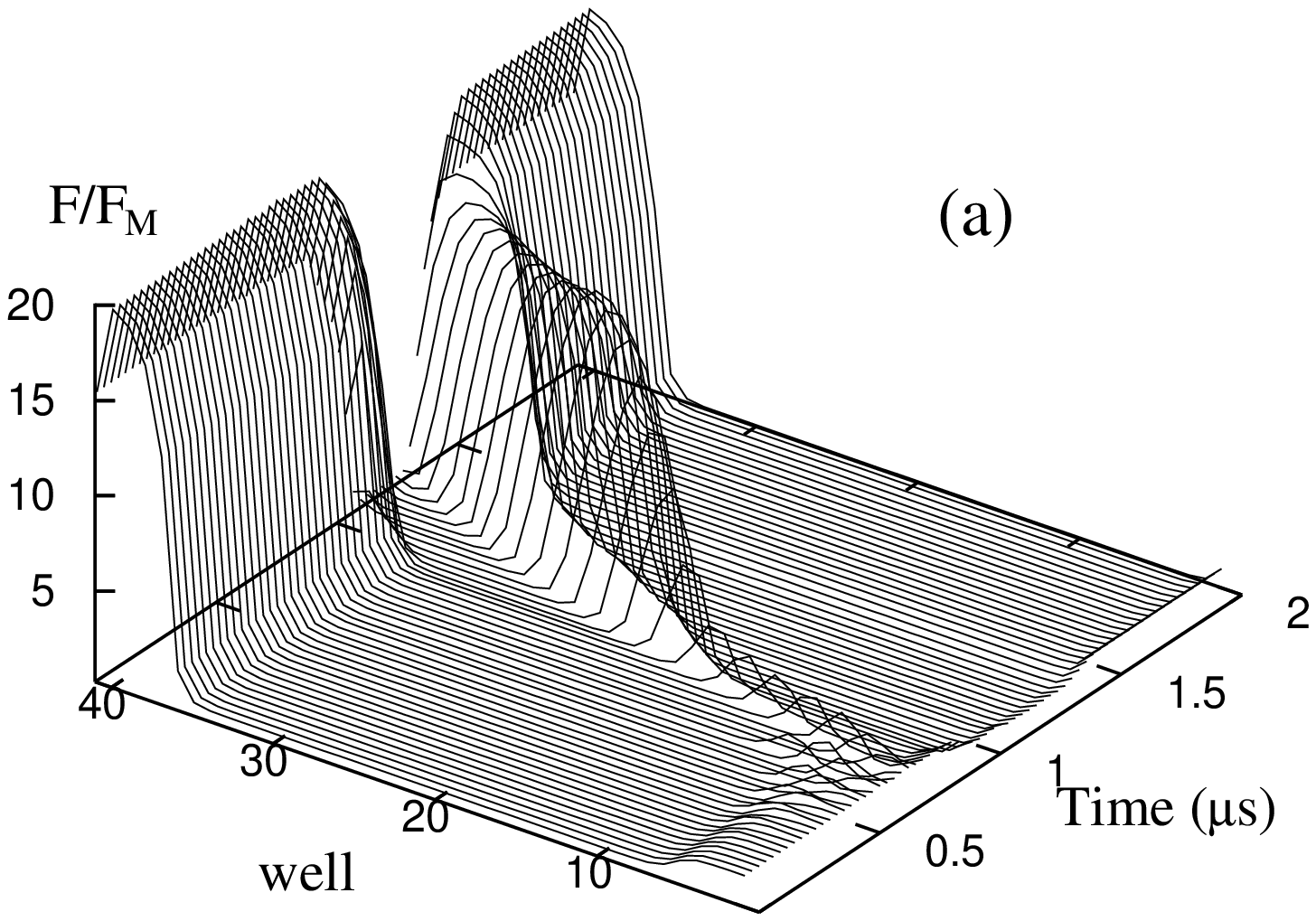} \hspace{-0.7cm}
\epsfxsize=90mm
\epsfbox{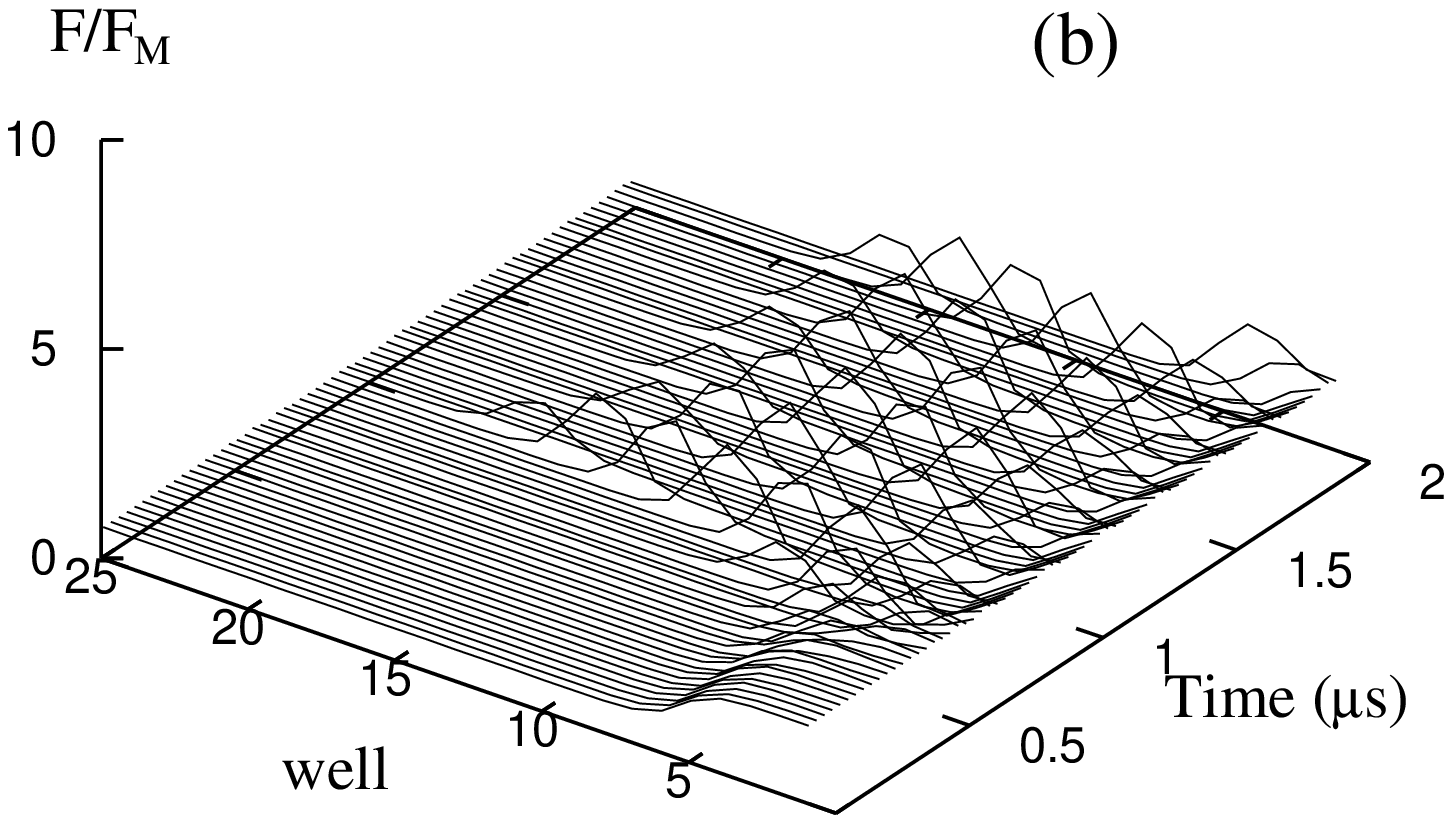}
}}
\caption{Evolution of the electric field profile corresponding to Fig.~\ref{fig12}(b), where
$V_i = 0.56$ V and $V_f = 0.59203$ V $<V_{th}$, for the two (very similar) ramping times
(a) $\tau_{r}= 2.1994\, \mu$s and (b) $\tau_{r}= 2.1995\, \mu$s. In (a), the CAL is emitted
after a short oscillatory transient. In (b), a CAL is emitted from the cathode, it disappears in 
the interior of the sample and this behavior is repeated periodically. This is similar to Gunn 
effect oscillations confined to one part of the SL. }
\label{fig13}
\end{center}
\end{figure}

\begin{figure}[ht]
\centerline{\hbox{
\epsfxsize=80mm
\epsfbox{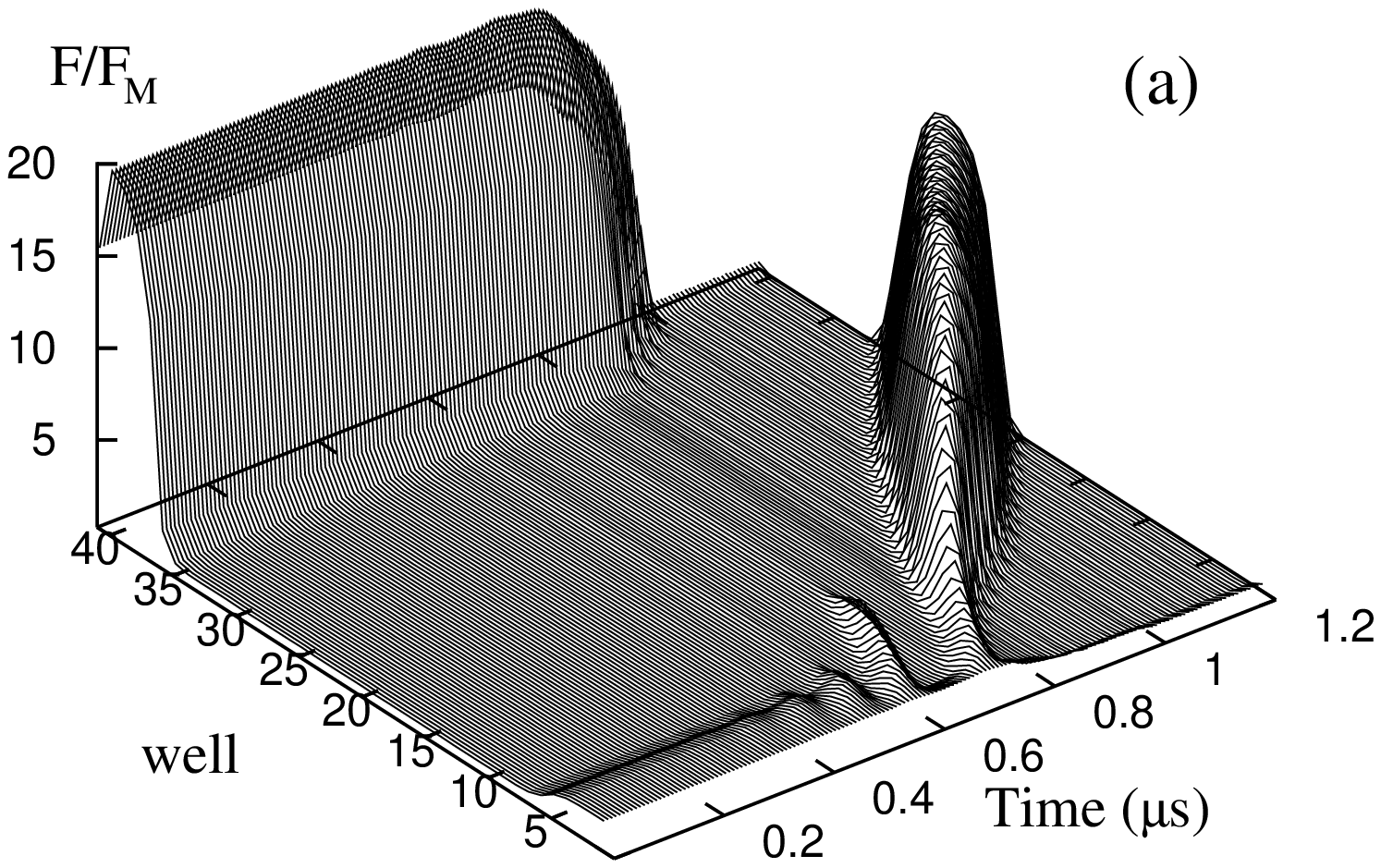} \hspace{0.7cm}
\epsfxsize=80mm
\epsfbox{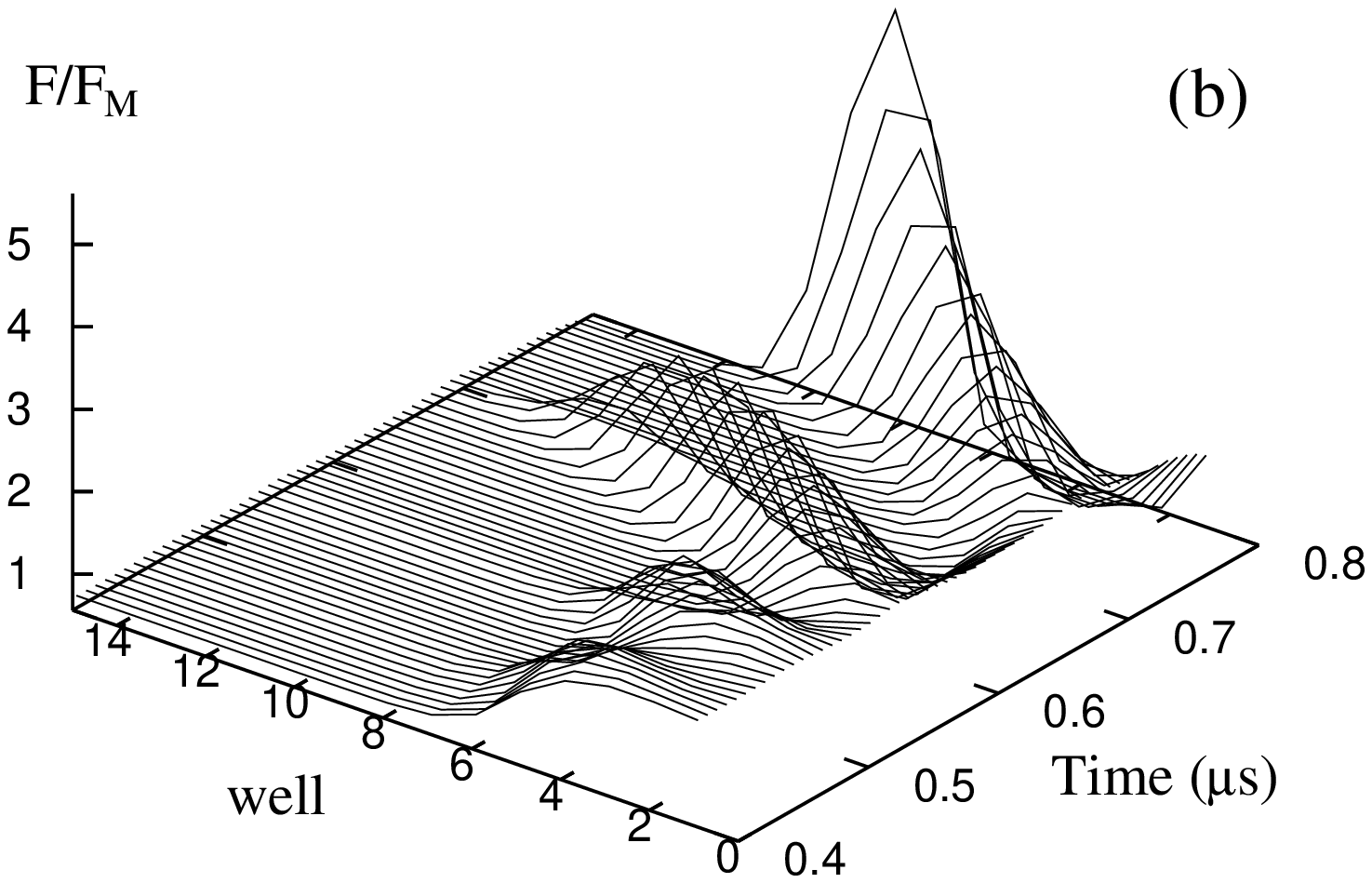}
}}
\caption{Evolution of the electric field profile corresponding to curve number 6 in
Fig.~\ref{fig12}(a), which has  longest oscillatory interval before a CAL is emitted 
from the cathode (ramping time $\tau_{r}= 20 \mu$s). (a) Detail of the oscillatory 
transient regime. (b) Detail of the profile near the cathode (closest 15 wells) when the CAL 
is finally emitted.}
\label{fig14}
\end{figure}

To ascertain the origin of the non-monotonic behavior of $t_{d}$ as a function of the 
ramping time, we observe that, as the voltage increases with time, the current becomes 
oscillatory before the tripole-dipole scenario begins. The shape of the current oscillations and 
their local period also change as the time elapses, which is clearly seen in Figs.~\ref{fig11} 
and \ref{fig12}(b). The latter figure also shows that, for similar ramping times, the current 
may drop to the lower I--V branch or continue oscillating for a longer time. Fig.~\ref{fig13}
depicts the field profiles during the oscillations of the current shown in Fig.~\ref{fig12}(b)
for $V_{f}<V_{th}$. We see that the current oscillations correspond to the periodic 
formation of a small field pulse at the cathode and its advance towards the anode over a few 
SL periods before it shrinks and vanishes. Eventually as the voltage increases with time, a 
dipole succeeds in growing sufficiently to detach itself from the cathode region and trigger a 
tripole-dipole event, bringing down the current to its stable value in the next I--V branch. 
Fig.~\ref{fig14} shows that the same mechanism is responsible for similar small amplitude
current oscillations for  $V_{f}>V_{th}$.

\begin{figure}[ht]
\centerline{\hbox{
\epsfxsize=80mm
\epsfbox{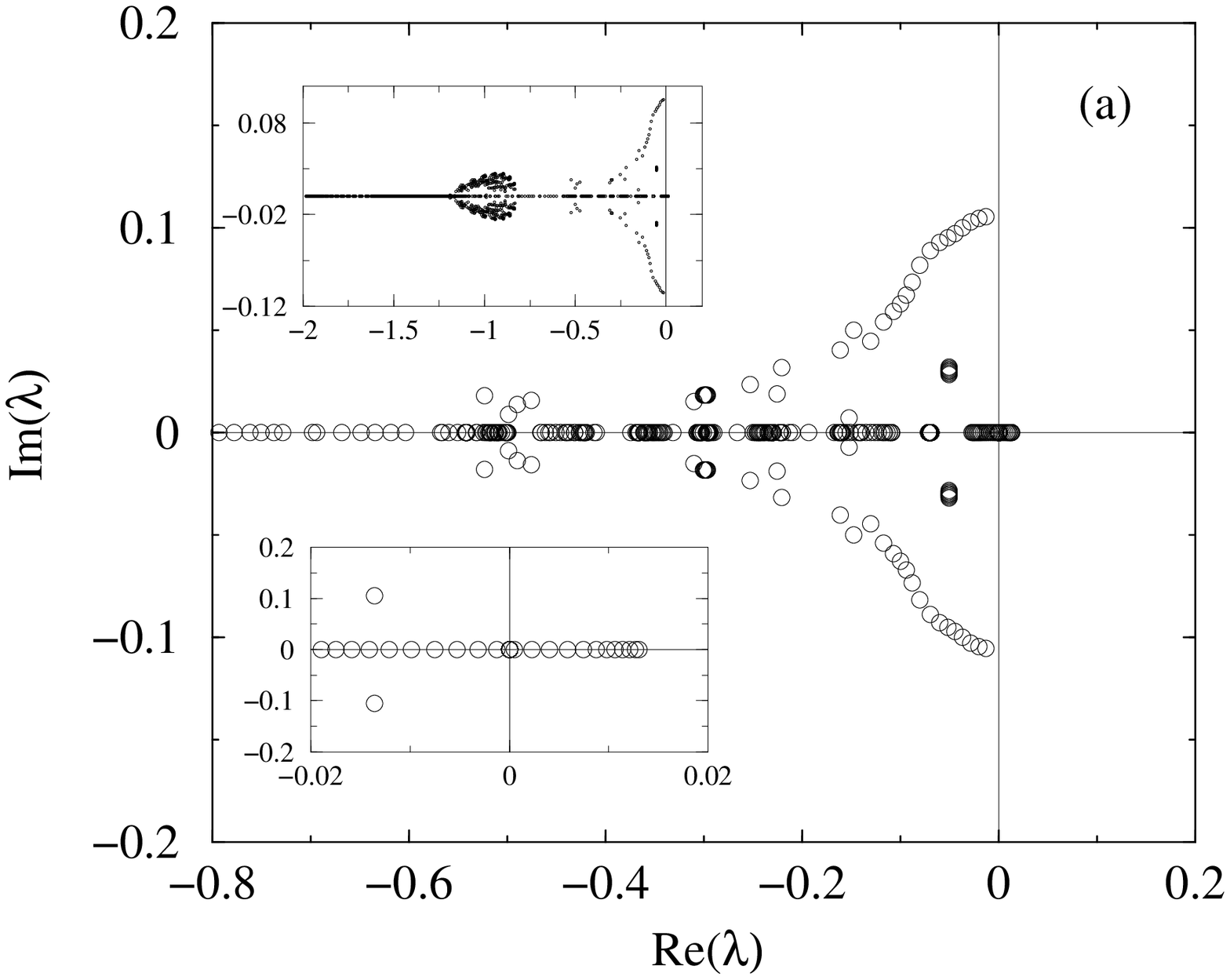} \hspace{0.7cm}
\epsfxsize=80mm
\epsfbox{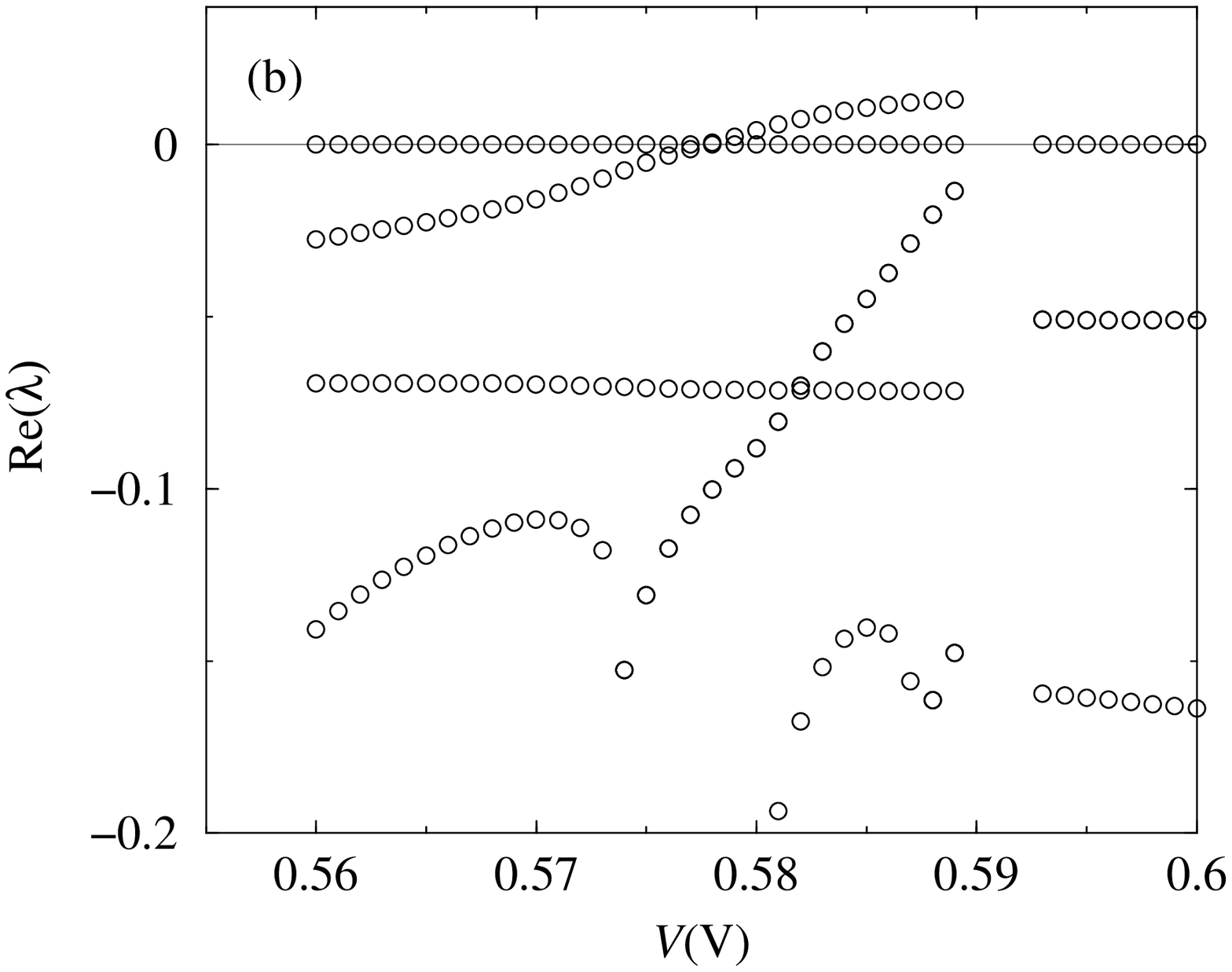}
}}
\caption{(a) Evolution of the 41 eigenvalues determining linear stability of the static 
Branch $B_{5}$ for 20 values of the applied voltage from $V=0.56$ to $0.6$ V. The insets
show the overall picture with all the eigenvalues and a zoom of the region near the imaginary
axis. (b) Real part of the eigenvalues as a function of voltage. The gap in the figure 
corresponds to the oscillatory instability.}
\label{fig15}
\end{figure}

Another clue to the different current response to switching at low voltage values is offered
by linear stability analysis of the I--V static solution branches. At low voltages, they have a 
stationary instability at a voltage smaller than $V_{th}$. At that voltage, a stationary branch 
bifurcates and this secondary branch undergoes a Hopf bifurcation to a small amplitude 
oscillation. The field profiles for these oscillatory solution correspond to the periodic emission
and motion of a small field pulse (charge dipole) confined to the region near the cathode, 
which is very similar to the confined Gunn effect in bulk n-GaAs  and ultrapure p-Ge 
\cite{BHi95}. Figures \ref{fig15}(a) and (b) show that the eigenvalue determining the 
linear stability of the static branch $B_{5}$ is real and it vanishes at a certain voltage smaller 
than $V_{th}$. The other eigenvalues with large real parts are complex and have negative real
parts. The gap in Fig.~\ref{fig15}(b) corresponds to the oscillatory instability observed
in the numerical simulations of the discrete model. It is plausible that the stationary branch 
that bifurcated from $B_{5}$ has a secondary oscillatory instability at that voltage, but 
a more detailed study is necessary before this can be ascertained. Thus we may have the 
following succession of bifurcations from the static branch $B_{5}$ as the voltage increases
towards $V_{th}$: (i) stable $B_{5}$, (ii) small amplitude static branch issuing from $B_{5}$,
(iii) Hopf bifurcation from the bifurcating static branch, (iv) annihilation of the oscillation
before or at $V_{th}$ (the end of $B_{5}$). During voltage sweeping, the current should
go through this succession of bifurcations and there is no reason why the time $t_{d}$ should
be a monotone function of the ramping time $\tau_{r}$.

\section{Conclusion}
\label{sec:7}
The relocation of the domain boundary in weakly coupled doped SL is substantially affected
by the ramping time over which the voltage is switched and by multistability of the initial 
and final static I--V branches involved in switching. Let us consider voltage switching leaving
several I--V branches between the initial and final voltages. If the ramping time is very long, 
the current simply follows adiabatically the change in voltage during switching, much as in up 
and down voltage sweeping. If the ramping time is very short, each branch jump during 
switching is achieved by a modified tripole-dipole scenario: a CDL is formed at the cathode, 
it moves towards the anode producing a second CAL behind it. Together with the old CAL,
the resulting charge tripole moves towards the anode until the first CAL and the CDL reach
it. Then the remaining CAL moves to its final position corresponding to the new static I--V
branch. For intermediate ramping time,  and provided the I--V branches have wide intervals 
of bistability, the tripole-dipole scenario may be skipped, thereby occurring every other 
branch jump. If the final voltage after switching is very close to the end of a I--V branch,
the current eventually drops to its value at the following static branch, but it can remain a 
long time on the initial static branch (or it oscillates about it in case there is an oscillatory 
instability) if the ramping time is sufficiently long. The time at which the tripole-dipole
scenario begins and the current drops to its stable value at the next I--V branch is not a
monotone function of the ramping time.
 
\section*{Acknowledgements}
This work has been supported by the MCyT grant MAT2005-05730-C02-01. R.E. has been 
supported by a postdoctoral grant awarded by the Autonomous Region of Madrid.

\appendix
\section{Linear stability of the static I--V branches}
\label{appA}
Let $\big(\{\mathcal{F}_i^*\}_{i=0}^N, \mathcal{J}^*\big)$ a stationary solution 
of (\ref{eqAdim}) - (\ref{JNAdim}) under dc voltage bias $\phi(\tilde{t})\equiv 
\phi$, $\forall \tilde{t}$. In these equations, we shall eliminate the electron density in 
favor of the field by using the Poisson equation. Then the tunneling current is a function of 
the electric field profile such that
\begin{eqnarray}
\mathcal{J}_{i \to i+1}(\{\mathcal{F}_i^*\}) = \mathcal{J}^*, \quad 
i=0,\dots,N,\\
\sum_{i=0}^N \mathcal{F}_i^* = (N+1) \phi.
\end{eqnarray}
Let $\big(\{f_i(\tilde{t})\}_{i=0}^N,j(\tilde{t})\big)$ a disturbance from the static 
solution:
\begin{eqnarray}
\mathcal{F}_i(\tilde{t}) = \mathcal{F}_i^* + \epsilon \, f_i(\tilde{t}), \quad 
\mathcal{J}(\tilde{t}) = \mathcal{J}^* + \epsilon \, j(\tilde{t}).
\end{eqnarray}
Then, the linear  equations about the static field profile and the static current density are
\begin{eqnarray}
{df_i\over d\tilde{t}} = j(\tilde{t}) - f_i \left. {\partial\mathcal{J}_{i \to i+1}
\over \partial \mathcal{F}_i} \right|_{(*)} - f_{i-1} \left. {\partial
\mathcal{J}_{i \to i+1}\over \partial \mathcal{F}_{i-1}} \right|_{(*)}
- f_{i+1} \left. {\partial\mathcal{J}_{i \to i+1}\over \partial \mathcal{F}_{i+1}} 
\right|_{(*)} , \quad i=0,\dots,N,
\end{eqnarray}
up to $O(\epsilon)$ terms. We have set $(*) = (\mathcal{F}_i^*,\mathcal{F}_{i-1}^*,
\mathcal{F}_{i+1}^*)$.

Let us now assume $f_i(\tilde{t})=e^{\lambda \tilde{t}} f_i$, $j(\tilde{t})=e^{
\lambda\tilde{t}} j$. Then we obtain
\begin{eqnarray}
\lambda f_i = j - f_i \left. {\partial\mathcal{J}_{i \to i+1}\over \partial
\mathcal{F}_i} \right|_{(*)} - f_{i-1} \left. {\partial\mathcal{J}_{i \to i+1}
\over \partial\mathcal{F}_{i-1}} \right|_{(*)}
 - f_{i+1} \left. {\partial\mathcal{J}_{i \to i+1}\over\partial
 \mathcal{F}_{i+1}} \right|_{(*)},
\end{eqnarray}
which can be written in matrix form as $\lambda {\bf f} = {\bf j} - {\bf A}\cdot {\bf 
f}$, with $a_{i,l} = \left. {\partial\mathcal{J}_{i \to i+1}\over \partial
\mathcal{F}_l} \right|_{(*)}$:
\begin{eqnarray}
\lambda \left(
\begin{array}{c}
f_0 \\
f_1 \\
\vdots \\
f_N
\end{array}
\right) = \left(
\begin{array}{c}
j \\
j \\
\vdots \\
j
\end{array}
\right)  -  \left(
\begin{array}{ccccc}
a_{0,0} & a_{0,1} & & & 0 \\
a_{1,0} & a_{1,1} & a_{1,2} & & \\
        & a_{2,1} & a_{2,2} & & \\
 & & & \ddots & a_{N-1,N} \\
 0 & & & a_{N,N-1} & a_{N,N}
\end{array}
\right) \cdot \left(
\begin{array}{c}
f_0 \\
f_1 \\
\vdots \\
f_N
\end{array}
\right) . \label{sistmat}
\end{eqnarray}
The boundary conditions for $i=0$ and $i=N$ yield $a_{0,0} = \tilde{\sigma}$, $a_{0,1} 
= 0$, $a_{N,N-1} = - \tilde{\sigma }\mathcal{F}^*_N/ \nu$ and $a_{N,N} = \tilde{
\sigma} (\tilde{n}^*_N + \mathcal{F}^*_N/\nu)$. On the other hand, the bias 
condition (\ref{biasAdim}) becomes
\begin{eqnarray}
\sum_{i=0}^N f_i = 0.
\label{biasei}
\end{eqnarray}
Then $\lambda {\bf f} + {\bf A}\cdot {\bf f} = {\bf j}$ implies
${\bf f} = (\lambda {\bf I} + {\bf A})^{-1} \cdot {\bf j}$. (\ref{biasei}) means that
the sum of the entries of the vector ${\bf f}$ is zero, therefore we have
\begin{eqnarray}
\sum_{i=0}^N  (\lambda {\bf I} + {\bf A})^{-1} \cdot {\bf 1}= 0, \label{poly}
\end{eqnarray}
because $j\neq 0$. The left hand side of (\ref{poly}) is polynomial of degree $N$ in
$\lambda$, having therefore $N$ zeros. For computational purposes, it is better to rewrite
the system (\ref{sistmat})-(\ref{biasei}) in the form $\lambda {\bf f} = {\bf B}\cdot 
{\bf f}$. This can be achieved adding the rows in (\ref{sistmat}) and using (\ref{biasei}):
\begin{eqnarray}
\lambda \sum_{i=0}^N f_i = (N+1)\, j - (a_{0,0}+a_{1,0}) f_0
- (a_{0,1}+a_{1,1}+a_{2,1}) f_1 - \dots \qquad \\
\dots - (a_{i-1,i}+a_{i,i}+a_{i+1,i}) f_i - \dots - (a_{N-1,N}+a_{N,N}) f_N = 0.
\end{eqnarray}
Defining $s_i = a_{i-1,i} + a_{i,i} + a_{i+1,i}$, $i=1,\dots,N-1$,
$s_0 = a_{0,0} + a_{1,0}$ and $s_N = a_{N-1,N} + a_{N,N}$, we have
\begin{eqnarray}
j = {1 \over N+1} \sum_{i=0}^N s_i f_i,
\end{eqnarray}
and therefore
\begin{eqnarray}
\left(
\begin{array}{c}
j \\
j \\
\vdots \\
j
\end{array}
\right) = {1 \over N+1} \left(
\begin{array}{ccccc}
s_0 & s_1 & s_2 & \dots & s_N \\
s_0 & s_1 & s_2 & \dots & s_N \\
\vdots &  \vdots &  \vdots &  &  \vdots \\
s_0 & s_1 & s_2 & \dots & s_N
\end{array}
\right) \cdot \left(
\begin{array}{c}
f_0 \\
f_1 \\
\vdots \\
f_N
\end{array}
\right) ,
\end{eqnarray}
which is of the form ${\bf j} = {1 \over N+1} {\bf S}\cdot{\bf f}$. Substituting
this expression in (\ref{sistmat}), we obtain a matrix equation of the type $\lambda {\bf f} 
= {\bf B}\cdot {\bf f}$, namely
\begin{eqnarray}
\lambda \left(
\begin{array}{c}
f_0 \\
f_1 \\
\vdots \\
f_N
\end{array}
\right) = \left\{
{1 \over N+1} \left(
\begin{array}{ccccc}
s_0 & s_1 & s_2 & \dots & s_N \\
s_0 & s_1 & s_2 & \dots & s_N \\
\vdots &  \vdots &  \vdots &  &  \vdots \\
s_0 & s_1 & s_2 & \dots & s_N
\end{array}
\right) - \left(
\begin{array}{ccccc}
a_{0,0} & a_{0,1} & & & 0 \\
a_{1,0} & a_{1,1} & a_{1,2} & & \\
 & & & \ddots & a_{N-1,N} \\
 0 & & & a_{N,N-1} & a_{N,N}
\end{array}
\right) \right\} \cdot \left(
\begin{array}{c}
f_0 \\
f_1 \\
\vdots \\
f_N
\end{array}
\right) 
\end{eqnarray}
The matrix ${\bf B}$ is equal to the matrix ${\bf S}/(N+1)$, except in its three main
diagonals, where $b_{i,j} = s_j/(N+1) - a_{i,j}$:
\begin{eqnarray}
\lambda \left(
\begin{array}{c}
f_0 \\
f_1 \\
\vdots \\
f_N
\end{array}
\right) = \left(
\begin{array}{ccccc}
b_{0,0} & b_{0,1} & b_{0,2} & \dots & b_{0,N} \\
b_{1,0} & b_{1,1} & b_{1,2} & \dots & b_{1,N} \\
\vdots & \vdots & \vdots & \ddots & \vdots  \\
b_{N,0} & b_{N,1} & b_{N,2} & \dots & b_{N,N}
\end{array}
\right) \cdot \left(
\begin{array}{c}
f_0 \\
f_1 \\
\vdots \\
f_N
\end{array}
\right) . \label{sistmat2}
\end{eqnarray}
The $(N+1)\times (N+1)$ matrix ${\bf B}$ has a zero eigenvalue (add its rows), and its 
other eigenvalues are the zeros of the polynomial (\ref{poly}). They have been depicted in
Fig.~\ref{fig10}.

\end{document}